\documentclass{elsarticle}

\usepackage[english]{babel}
\usepackage{lineno,hyperref}
\usepackage{units}
\usepackage{graphicx}
\usepackage{amsmath}
\usepackage{amssymb}
\usepackage{subfigure}

\begin{document}
\begin{frontmatter}

\title{A parameterization for the radio emission of air showers as predicted by CoREAS simulations and applied to LOFAR measurements}

\author[ru,ni]{Anna Nelles}
\cortext[mycorrespondingauthor]{Corresponding author}
\ead{a.nelles@astro.ru.nl}
\author[ru]{Stijn Buitink}
\author[ru,ni,as,mpfir]{Heino Falcke}
\author[ru,ni]{J\"org R. H\"orandel}
\author[kit]{Tim Huege}
\author[ru]{ Pim Schellart}

\address[ru]{Department of Astrophysics/IMAPP, Radboud University Nijmegen, 6500 GL Nijmegen, The Netherlands}
\address[ni]{Nikhef, Science Park Amsterdam, 1098 XG Amsterdam, The Netherlands}
\address[as]{Netherlands Institute for Radio Astronomy (ASTRON), 7990 AA Dwingeloo, The Netherlands}
\address[mpfir]{Max-Planck-Institut f\"ur Radioastronomie, Auf dem H\"ugel 69, 53121 Bonn, Germany}
\address[kit]{Institut  fŸr  Kernphysik,  Karlsruhe  Institute of Technology (KIT), Germany}

\begin{abstract}
Measuring radio emission from air showers provides excellent opportunities to directly measure all air shower properties, including the shower development. To exploit this in large-scale experiments, a simple and analytic parameterization of the distribution of the radio signal at ground level is needed. Data taken with the Low-Frequency Array (LOFAR) show a complex two-dimensional pattern of pulse powers, which is sensitive to the shower geometry. Earlier parameterizations of the lateral signal distribution have proven insufficient to describe these data. In this article, we present a parameterization derived from air-shower simulations. We are able to fit the two-dimensional distribution with a double Gaussian, requiring five fit parameters. All parameters show strong correlations with air shower properties, such as the energy of the shower, the arrival direction, and the shower maximum. We successfully apply the parameterization to data taken with LOFAR and discuss implications for air shower experiments. 
\end{abstract}

\begin{keyword}
cosmic rays\sep  air shower\sep radio emission\sep lateral distribution \sep LOFAR 
\end{keyword}

\end{frontmatter}

\section{Introduction}
Radio emission of air showers received a lot of attention in the 1960s \cite{Allan1971}. The emission was discovered by Jelley et al. in 1966 \cite{Jelley1965} and soon a number of theorists and experimentalists tried to explain and study the emission. Whereas it is today widely agreed upon that the emission at MHz frequencies is a mix of emission created by the deflection of relativistic electrons and positrons in the geomagnetic field (geomagnetic effect) \cite{Kahn1966, Falcke2003, Huege2003} and the charge separation along the shower axis (charge excess) \cite{Askaryan1962, James2011, Huege2013}, this was an unresolved discussion in the 1960s. 

Today, our theoretical understanding of the emission is based on a number of different air shower simulations and models. Different strategies are used to describe air showers and their emission: some models are embedded in full Monte Carlo simulations, tracking every individual particle and calculating its emission (CoREAS \cite{Huege2013b}, ZHAireS \cite{Alvarez-Muniz2011}), while others use shower universality to consequently calculate the emission (SELFAS2 \cite{Marin2012}) or apply a macroscopic description of the emission to a parameterized air shower (EVA \cite{Werner2012}).
 The development of many models was accompanied by early results from air shower experiments such as LOPES \cite{Falcke2005} or CODALEMA \cite{Ardouin2005}, which aided the understanding of the emission and relevant parameters. All of the models do not only follow the development of the air shower, but also include detailed models of the refractive index of the atmosphere, which significantly influences the detected emission \cite{Vries2010}. None of the models, however, provides an analytic parameterization of the signal distribution at ground level. 

Radio emission has been pursued as a detection technique as it is directly sensitive to the shower development. The achievable resolution of the shower maximum is comparable to fluorescence detectors \cite{Apel2012, Stijn, Palmieri2013}, while allowing for much longer duty-cycles. As the origin of the cosmic rays at the highest energies is still unknown, experiments with the ability to collect large event statistics are needed, which provide at the same time a good resolution of energy and shower maximum and thereby the mass of the cosmic rays. 

Data taken with the Low-Frequency Array (LOFAR) \cite{Schellart2013}, have shown the sensitivity of the emission pattern to the shower maximum \cite{Stijn}, but have also made the need for a more complex model of the lateral signal distribution more visible. Due to its high density of antennas, LOFAR is the most suitable experiment to measure subtle features in the emission and to test models of the emission. 

In this article, we first review the current knowledge of the \emph{radio lateral distribution function (radio LDF)}, i.e. the pulse power or amplitude as function of distance to the shower axis (section \ref{theory}). This review is then followed by general considerations based on air shower simulations, which are used to develop a parameterization of the signal (sections \ref{sec:sim_set}, \ref{sec:gen}). The model obtained is applied to a large set of simulations to discuss the sensitivity towards shower parameters (sections \ref{sec:dis} and \ref{sec:phys}). Finally in section \ref{sec:data}, we present a reduced and more robust model, which is applied to LOFAR data and is able to convincingly reproduce the measurements. For LOFAR, this analytic parameterization will speed up the process of reconstructing $X_{\mathrm{max}}$, as it reduces the parameter space for detailed full air shower simulations, which currently require a significant amount of computation time. This parameterization will also be of importance for other air shower detectors measuring the radio emission. 

\section{Theoretical models and earlier parameterizations}
\label{theory}
Almost all currently available experimental data have been described, based on the early work and the thereby established parameterization by Allan. He argues in his extensive review of the early data and analysis \cite{Allan1971} along the following lines to derive his parameterization:

The observed amplitude of the electric field should be proportional to the sine of the angle between shower axis and magnetic field, $\alpha$, derived from its induced direction in $\vec{v}\times\vec{B}$, where $\vec{v}$ is the arrival direction of the cosmic ray and $\vec{B}$ the direction of the local geomagnetic field.  Furthermore, the amplitude should be proportional to the energy of the primary particle, although this is only claimed for energies between  $\unit[10^{17}]{eV}$ and  $\unit[10^{18}]{eV}$. At higher energies, he expects a steep increase of the signal amplitude, as the shower maximum comes closer to the observer.  Additionally, he expects the radial distribution to broaden with zenith angle due to geometric considerations, with at the same time a decrease in peak amplitude. This effect is predicted to be opposed by the increase of efficiency of the geomagnetic emission due to the decreased air density at the shower maximum. 

Together with experimental data, these predictions were summarized in the following equation for the radio pulse amplitude $E_{\nu}$ per unit bandwidth:
\begin{equation}
E_{\nu} = C \cdot \left(  \frac{E_p}{10^{17}\mathrm{eV}} \right) \sin(\alpha) \cos(\theta) e^{-\frac{R}{R_0(\nu,\theta)}} \unit{\frac{ \mu V }{m\ MHz}}.
\end{equation}
For values of $\unit[10^{17}]{eV} < E_p < \unit[10^{18}]{eV}$ for the energy of the primary particle and distances $R< \unit[300]{m}$, Allan gives the scaling parameter $C = 20$, $R_0 = 100 \pm \unit[10]{m}$ for $\nu=\unit[54]{MHz}$ and zenith angles $\theta < 35^{\circ}$.

This \emph{Allan-formula} was consequently used at all later experiments to describe the lateral distribution. The CODALEMA experiment \cite{Ardouin2005} used the same parametrization, albeit with a different scaling factor. It was found that when leaving the core position as a free parameter the \emph{radio core} (ground location of the highest radio amplitude) showed a significant offset with respect to the particle core \cite{Rebai2011, MarinICRC}, but that otherwise the measurements were fairly well represented. 

Also the LOPES (LOFAR PrototypE Station) experiment \cite{Falcke2005} used the same parameterization. While the Allan formula refers to the total electric field, the LOPES experiment first measured only one component (East-West), for which it was argued that the $\sin(\alpha)$ dependence might rather be a $1-\cos(\alpha)$-dependence \cite{Horneffer2007}. Also,  different slopes $R_0$ and scaling parameters $C$ were fitted. 

The challenge in comparing different scaling parameters resides in the complexity of obtaining an absolute calibration of the measured amplitude of the experiments. The earlier experiments used narrow-band receivers, rather simple antennas and oscilloscopes. The more modern experiments use broad-band systems with more complex antennas and analogue chains. To measure the characteristics of these set-ups with the necessary precision as a function of frequency is challenging and subject to a number of systematic uncertainties \cite{Nehls2008}. Additionally, there is a geographic component given the different strengths of the local geomagnetic field, as well as different heights above sea level of the different experiments. 

As the air shower models improved, more theoretical studies concerning the lateral distribution were conducted \cite{Huege2005}, showing dependences on the height of the shower maximum and that the Allan parameterization might be difficult to hold \cite{Vries2010, Alvarez-Muniz2011}. Especially, studies predicting asymmetries in the pattern, called for a more complex function. For LOPES, a fit of a one dimensional Gaussian was suggested, which was offset with respect to the shower axis. The slope or width of the Gaussian was found to be a function of the height of shower maximum \cite{Palmieri2013}. 

The experimental data of CODALEMA and LOPES, as well as the early measurements, were reasonably well described by the one-dimensional Allan-formula and adaptions of it, all preserving the exponential fall-off.  However, LOPES also observed some \emph{flat} lateral distributions, which could not be explained by the exponential parameterization \cite{Apel2010}. 

\begin{figure}
\begin{center}
\includegraphics[width=0.49\textwidth]{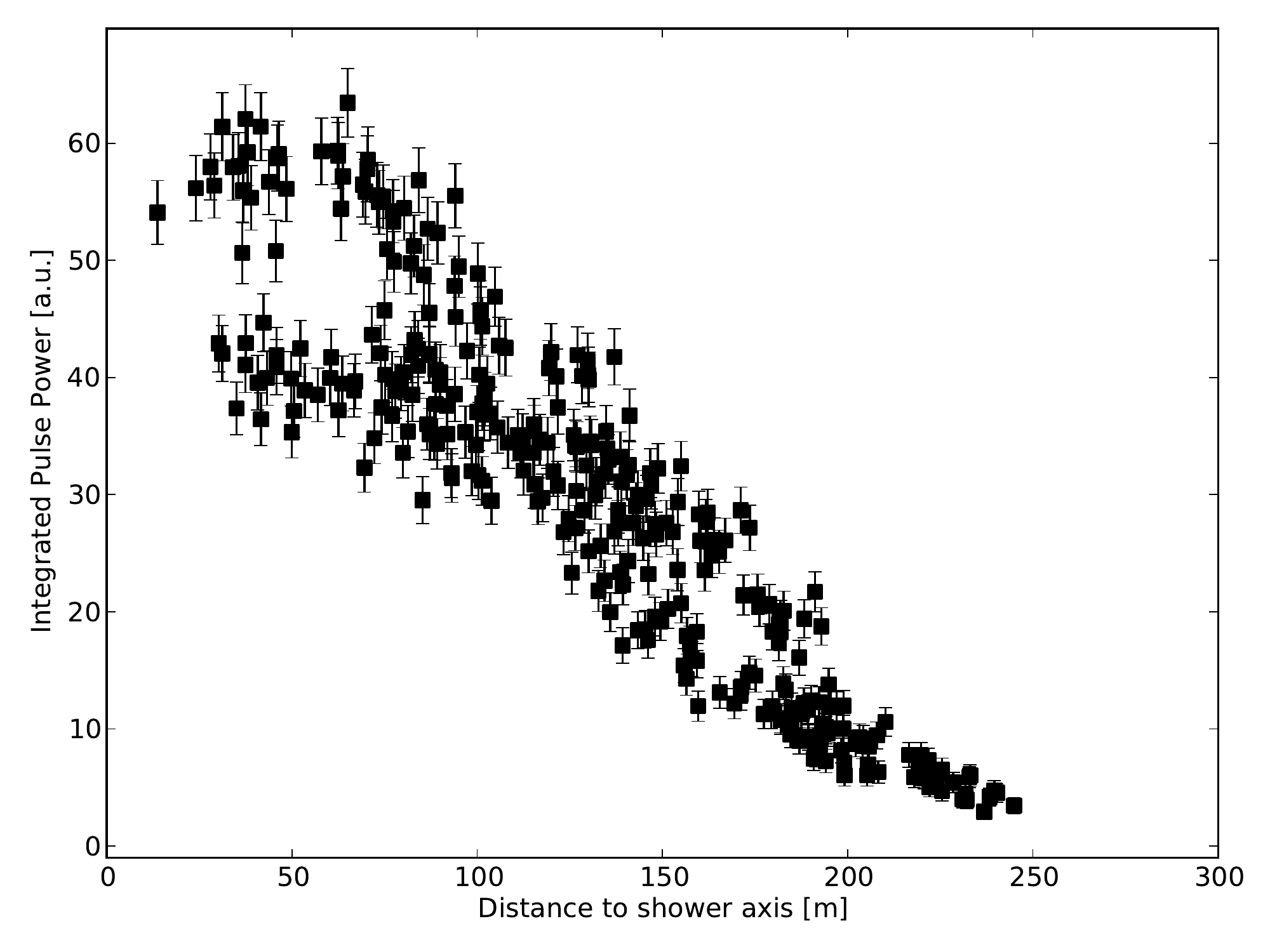}
\caption{Radio emission of an air shower, as measured with LOFAR. The arrival direction had a zenith angle of $\theta=30.8^{\circ}$ and and azimuth angle of $\phi=127.7^{\circ}$. As there is currently no absolute calibration of the LOFAR antennas, the integrated pulse power, calibrated relative for all antennas, is plotted as a function of the distance to the shower axis \cite{Schellart2013}. A number of detailed structural features are visible. These are due to measurements at the same distance from the shower axis, but at different azimuthal angles, which cannot be represented in this one-dimensional projection.}
\label{fig:ldf_data}
\end{center}
\end{figure}

Both LOPES and CODALEMA measured single air showers with a maximum of about 25 antennas per event on relatively small distance scales ($\unit[200]{m}$).  When LOFAR \cite{LOFAR} started taking data, with more than 500 antennas per event on scales covering more than $\unit[500]{m}$ it became obvious that a one-dimensional LDF was unable to describe the data, given significant asymmetries as can be seen in figure \ref{fig:ldf_data}.

Based on air shower simulations \cite{Huege2013a}, we derive a two-dimensional parameterization that is fitted to the simulations generated for LOFAR and subsequently to the data. 

\section{Air shower simulations}
\label{sec:sim_set}
A large set of air shower simulations is available to the LOFAR key science project Cosmic Rays. They were originally made to determine the depth of the shower maximum, $X_{\mathrm{max}}$, of every LOFAR shower by directly comparing the data to single simulations \cite{Stijn}. For every shower measured with a certain number of antennas with LOFAR, 40 simulated air showers with different $X_{\mathrm{max}}$-values (both proton and iron primaries) were generated, all matching the reconstructed arrival direction of the measured shower. The energy of a simulated shower, however, is not necessarily the energy of the corresponding measured shower. The value that was simulated is based on the energy estimated from the particle detectors installed at LOFAR. As many of the showers measured were contained in the radio array, but not in the particle array, the energy estimate may differ by a factor of ten to the actual shower energy. 

The air shower simulations are produced using CORSIKA 7.400 with FLUKA 2011.2b and QGSJETII.04 in the US standard atmosphere. A thinning of $10^{-6}$ is applied. The radio emission is generated by the CoREAS plug-in \cite{Huege2013b}. 

\begin{figure}
\begin{center}
\includegraphics[width=0.49\textwidth]{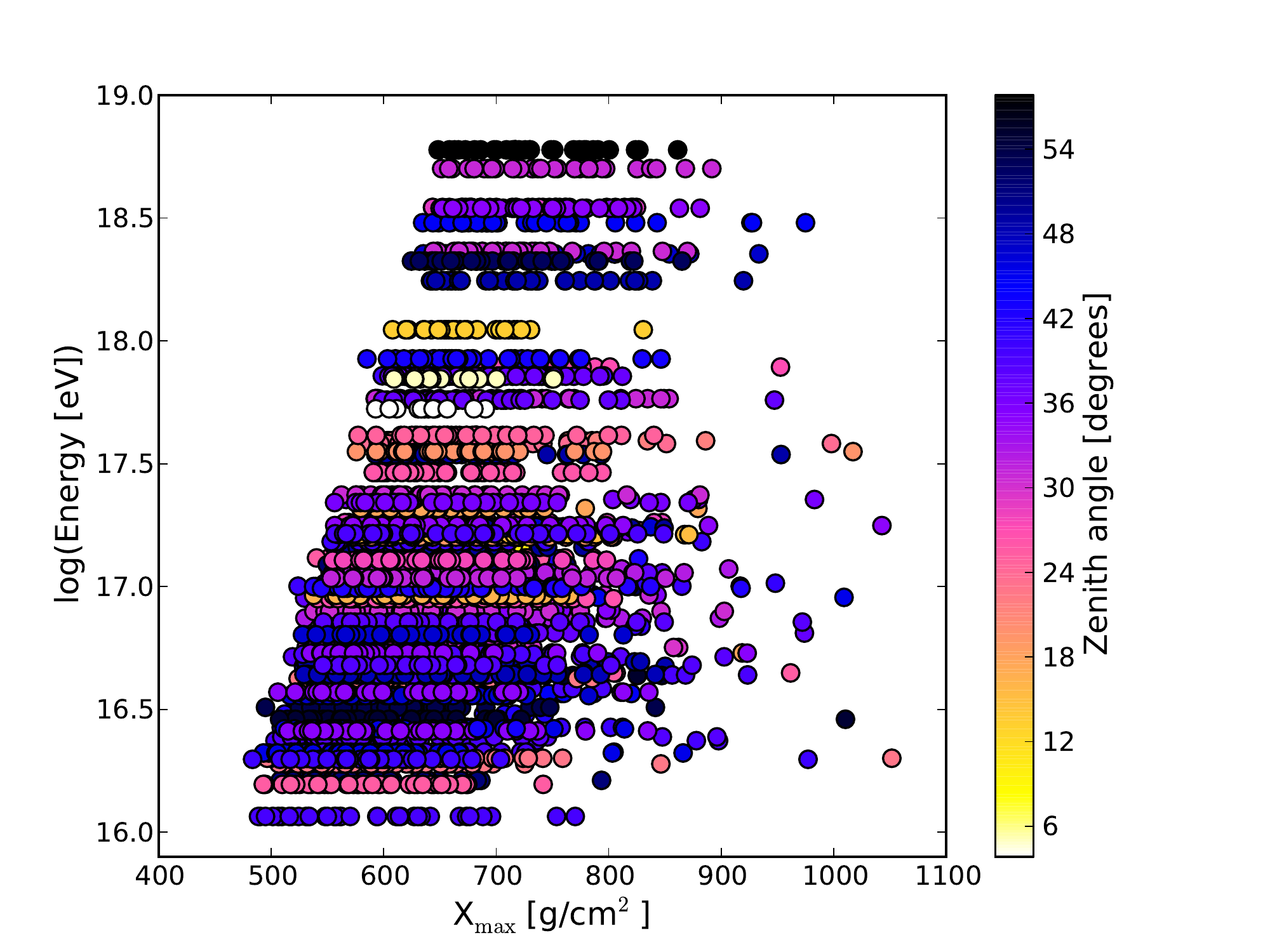}
\includegraphics[width=0.49\textwidth]{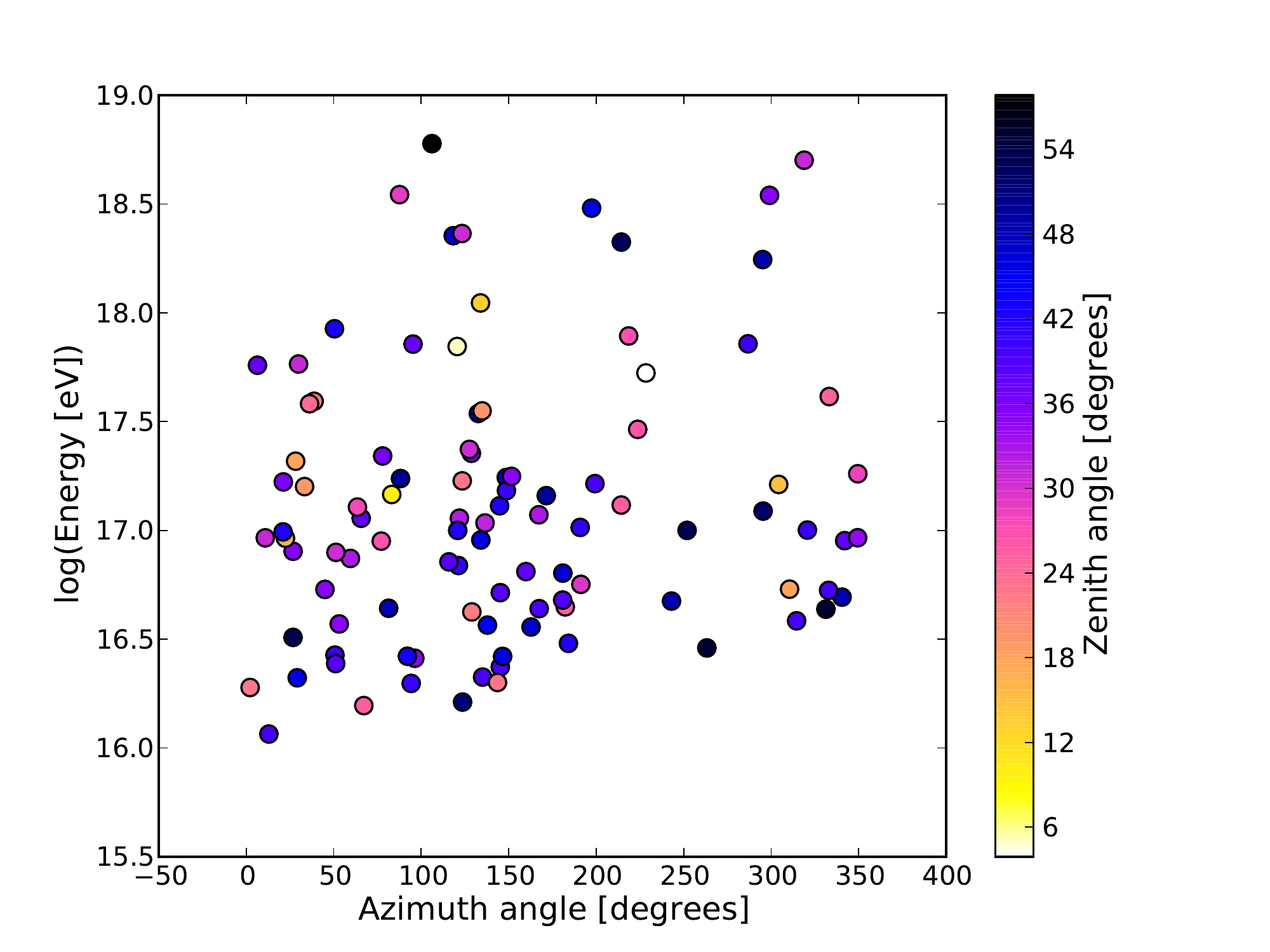}
\caption{Parameters of simulated air showers used in this analysis. Left: Energy of the simulated shower as a function of  the depth of the shower maximum, $X_{\mathrm{max}}$. Different energies of $\unit[10^{16}-10^{18.8}]{eV}$ are sampled, each one with 40 different values of $X_{\mathrm{max}}$. The higher energies are less densely sampled. The zenith angles ($0^{\circ}$: vertical) are encoded in color. No horizontal air showers ($> 60^{\circ}$) were simulated. Right: The distribution of arrival directions as a function of energy. No correlation between the parameters is visible. The azimuth angles ($90^{\circ}$ being north) are not uniformly sampled, there is a bias against showers arriving parallel to the local magnetic field. }
\label{fig:simpar}
\end{center}
\end{figure}

The distributions of all shower parameters are shown in figure \ref{fig:simpar}. The simulated showers span an energy range from $\unit[10^{16}-10^{18.8}]{eV}$  and cover zenith angles $\theta$ from $3^{\circ}$ to $55^{\circ}$, where $0^{\circ}$ are vertical showers. There is no correlation between the arrival direction and energy. For the overall distribution of angles, it should be noted that there are relatively few showers from a direction parallel to the magnetic field, as LOFAR is less likely to observe radio emission of showers from this direction \cite{Schellart2013}. The geomagnetic field is pointing directly North with an inclination angle of $67^{\circ}$ downwards.  This means that there are relatively few showers with azimuth angles $\phi$ close to south, i.e. $270^{\circ}$. 

\begin{figure}
\includegraphics[width=0.49\textwidth]{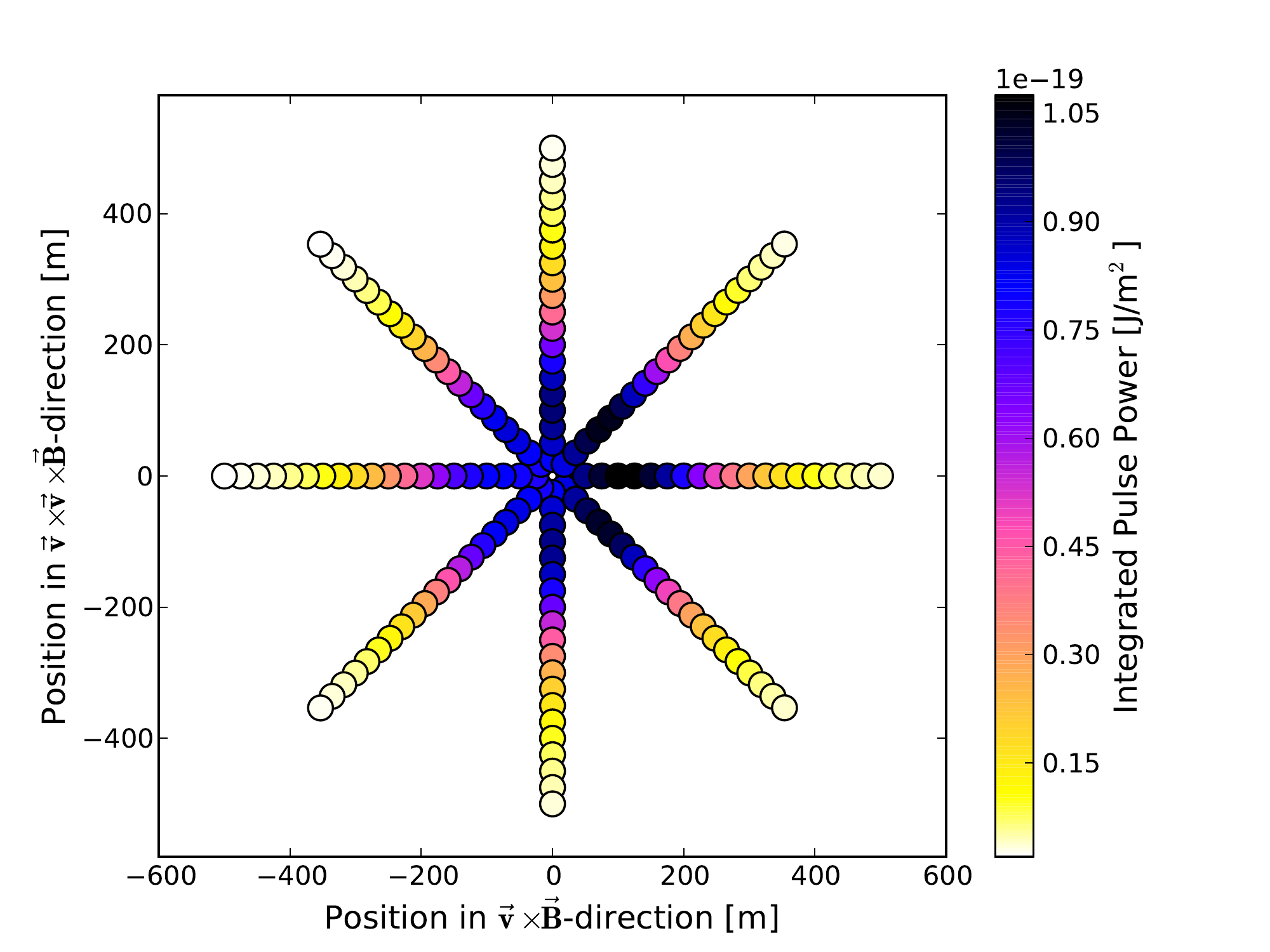}
\includegraphics[width=0.49\textwidth]{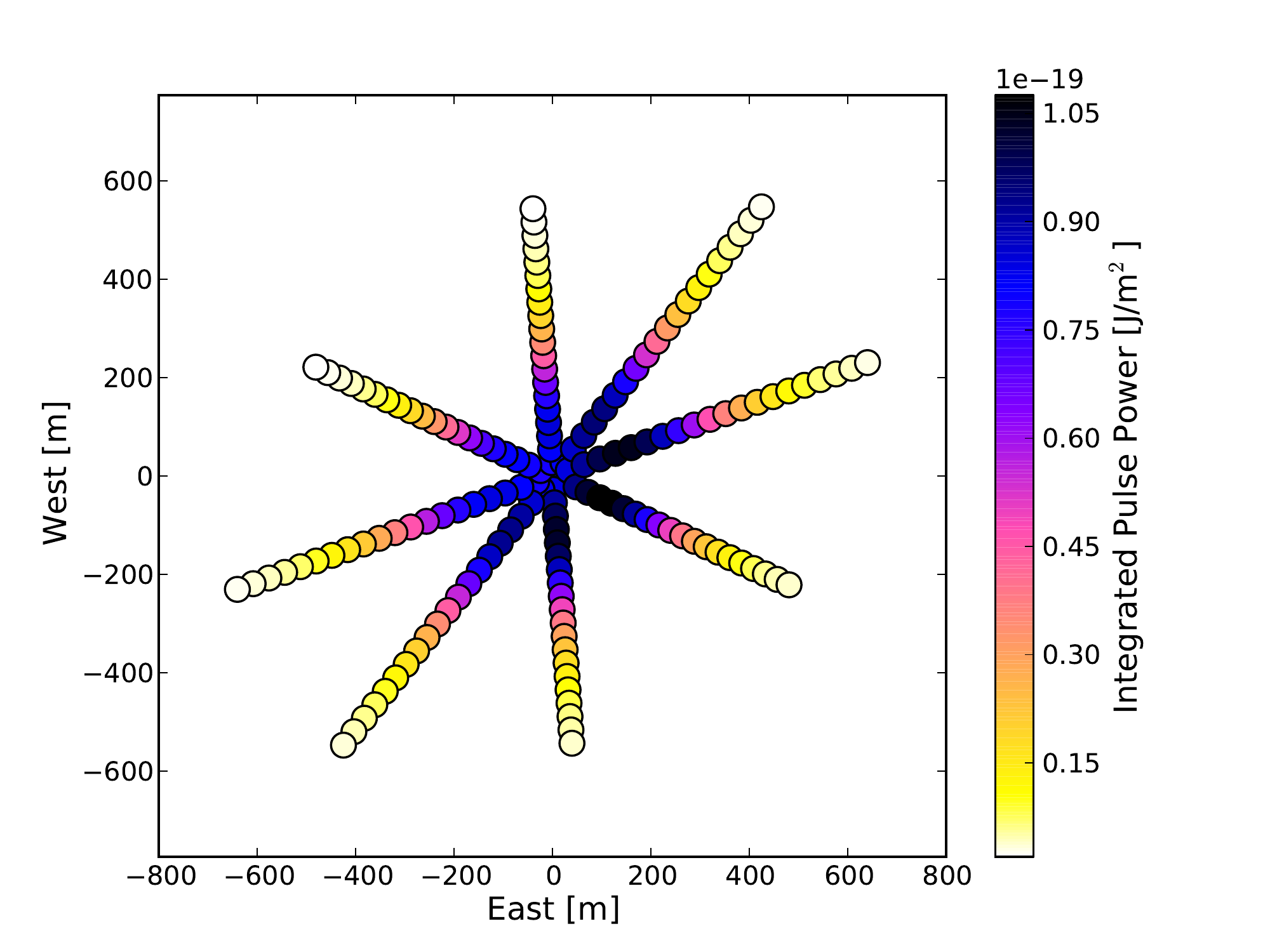}
\caption{Grid of antennas on which the air shower was simulated. The left side shows the antennas in shower coordinates ($\vec{v}$ is the direction of the shower axis, $\vec{B}$ the direction of the magnetic field) and the right side depicts the positions on the ground. The integrated power of the simulated pulses is encoded in color. This simulated air shower arrived under a zenith angle $\theta = 45^{\circ}$.}
\label{fig:grid}
\end{figure}

The simulations are performed on an idealized grid of antennas as shown in figure \ref{fig:grid}. The grid is aligned in such a way that it is always aligned with the $\vec{v} \times \vec{B}$-axis and the $\vec{v}\times\vec{v}\times\vec{B}$-axis, where $\vec{v}$ is the direction of the shower and $\vec{B}$ the direction of the magnetic field. It is therefore rotated and stretched differently on the ground plane for every shower. The ground plane at LOFAR is located $\unit[5]{m}$ above sea level. 

The CoREAS simulations deliver the resulting electric field per antenna position as a function of time, in this case at a resolution of $\unit[0.1]{ns}$. The simulations are subsequently downsampled to the LOFAR sampling frequency of $\unit[200]{MHz}$ and filtered from $\unit[10-90]{MHz}$, matching the LOFAR low-band antenna measurements. For every simulated antenna position, the signal in the time-domain is squared to obtain the power and added up, delivering the integrated power. This is calculated for every polarization and subsequently added up to receive the total power. This calculation is performed in the same way, as it is done to the data \cite{Schellart2013}. The integrated total signal is chosen for comparison as it is only affected by the absolute bandpass of the experiment and not sensitive to the frequency dependent phase response. Possible uncertainties in the modeling of the phase response of the system will average out for the integrated quantities for both the signal and the background contribution, while being a relevant factor for measurements of the pulse amplitude. Also, changes in the frequency spectrum of the pulses as a function of distance to the shower axis \cite{REASMGMR} will affect the form of the pulse and thereby its maximum amplitude, while preserving the power. By choosing the integrated power, the effect of  the change in frequency spectrum and the decreasing power are separated and only the latter is discussed in this analysis. 

\section{General considerations and choice of parametrization}
\label{sec:gen}
\begin{figure}
\begin{center}
\includegraphics[width=0.49\textwidth]{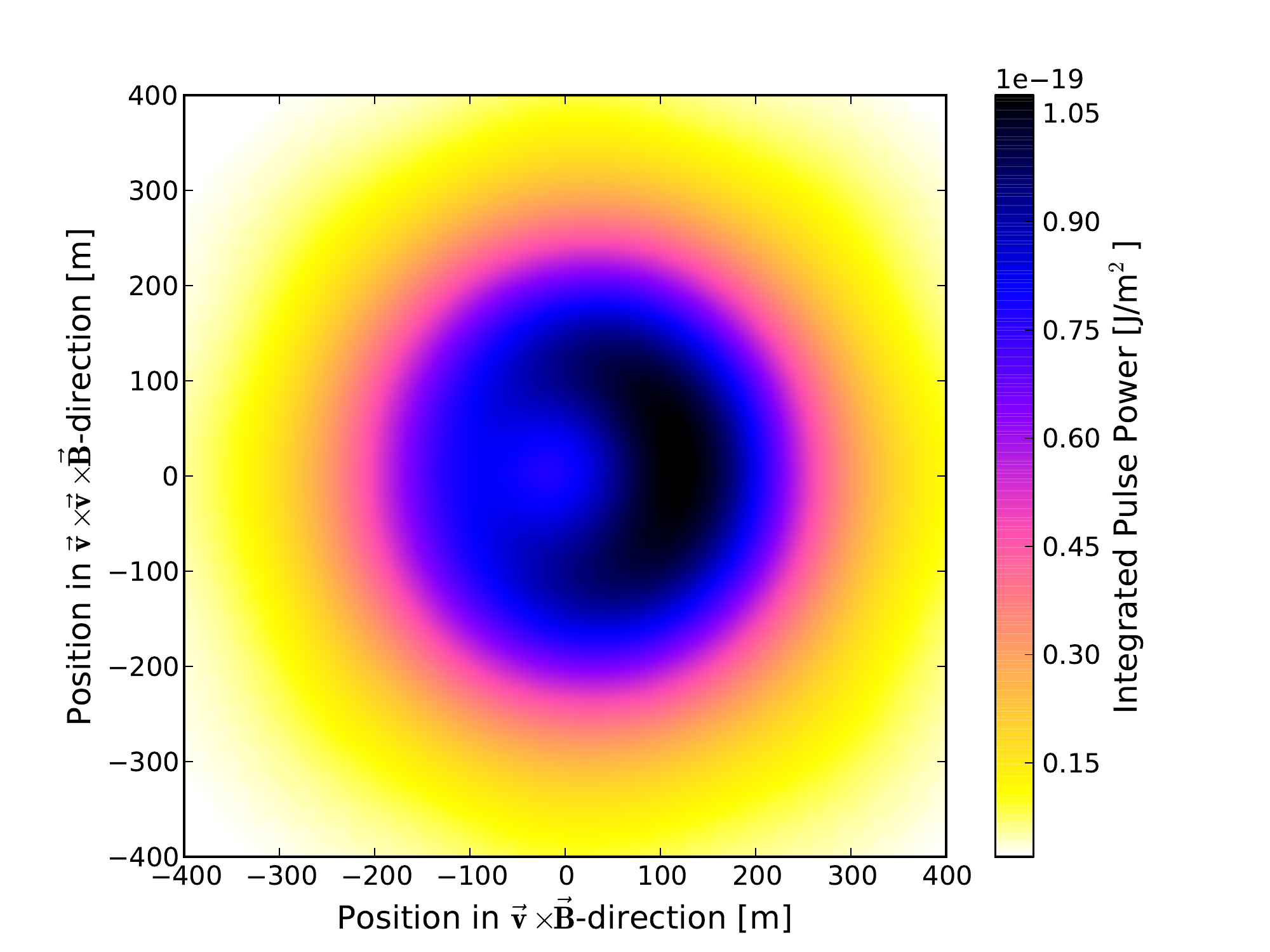}
\includegraphics[width=0.49\textwidth]{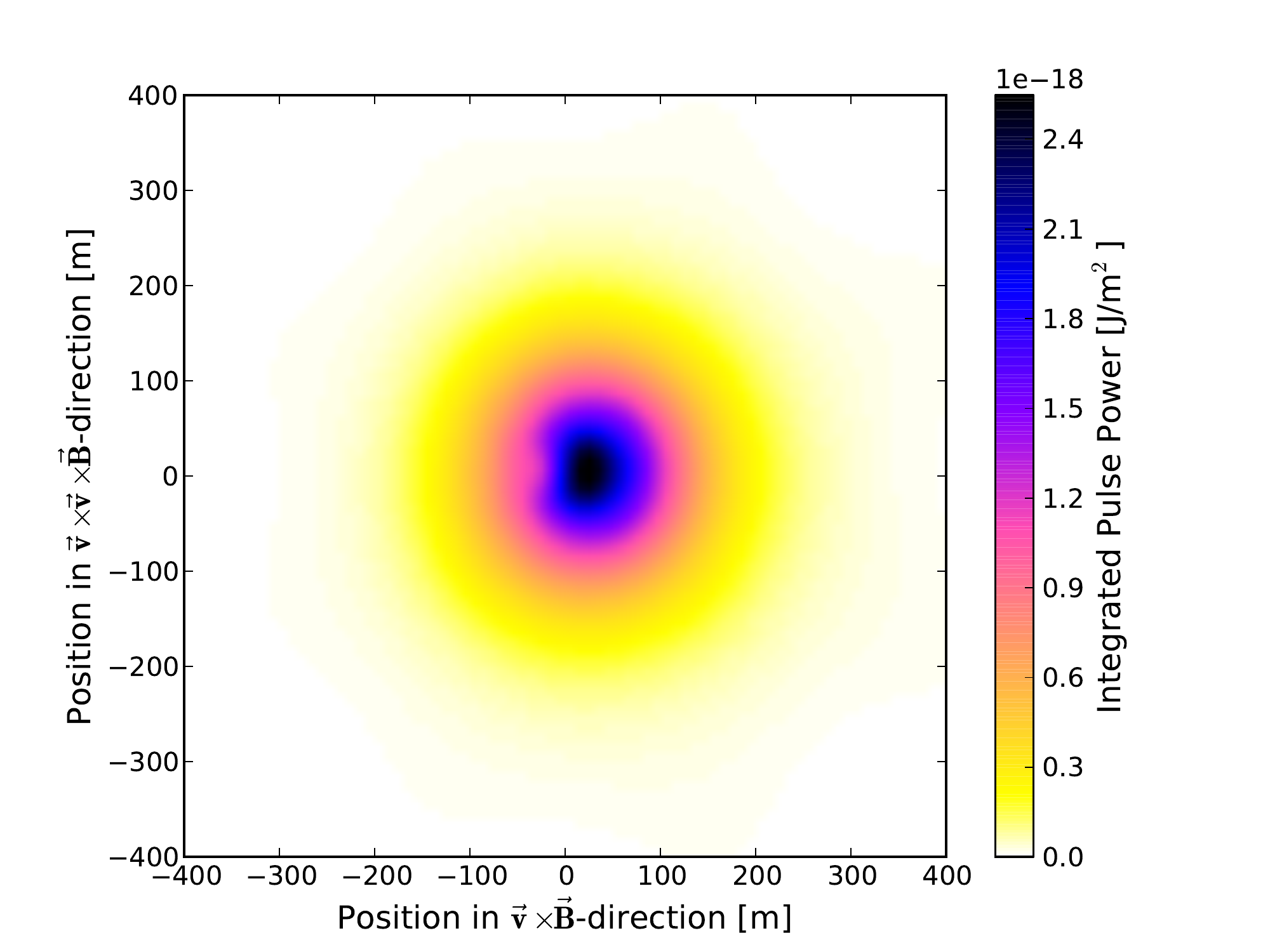}
\caption{Interpolated pattern of the simulated total power for two different air showers in the shower plane. Left shower: $\theta=45^{\circ}, \phi=37^{\circ}, \alpha=61^{\circ},  E= \unit[4.4\cdot 10^{16}]{eV},X_{\mathrm{max}}=\unit[640]{g/cm^2}$. Right  shower: $\theta=29^{\circ}, \phi=121^{\circ}, \alpha=49^{\circ}, E= \unit[1.1\cdot 10^{17}]{eV},X_{\mathrm{max}}=\unit[823]{g/cm^2}$. Despite being measured at largely different distances to the shower maximum, at different arrival directions, and at different energies, both showers show a visible asymmetry and a circular, bean-shaped pattern. }
\label{fig:pattern}
\end{center}
\end{figure}

In order to better visualize the shape of the lateral signal distribution of the simulated signal, the power from the grid pattern (figure \ref{fig:grid}) can be interpolated and plotted, as it is done in figure \ref{fig:pattern}. Since this is in the shower plane, this pattern is in general circular, so one is tempted to look for rotational symmetry. It is however also clearly visible that the central part with the highest signal is not rotationally symmetric. 

As discussed in section \ref{theory}, the classical choice is an exponential function. Especially for events measured at larger distances to the shower axis, this has proven to be successful. Thus, functions which have an exponential fall-off at larger distances are obvious candidates. In addition, the functions should deliver a flattening or even fall-off near the center.  Purely from these shape considerations, the following initial parameterization is chosen.

\begin{equation}
P(x^{\prime},y^{\prime}) = A_+ \cdot \exp\left(\frac{-[(x^{\prime}-X_+)^2+(y^{\prime}-Y_+)^2]}{\sigma_+^2}\right) - A_-\cdot  \exp\left(\frac{-[(x^{\prime}-X_-)^2+(y^{\prime}-Y_-)^2]}{\sigma_-^2}\right) + O
\label{eq:ini}
\end{equation}

Here, $P$ is the total power of the integrated radio signal (sum of the powers from all polarizations). The coordinates $x^{\prime},y^{\prime}$ are in the shower plane, centered around the position of the shower axis . The shower plane is spanned by the vectors $\vec{v} \times \vec{B}$ and $\vec{v}\times\vec{v}\times\vec{B}$, where $x^{\prime}$ and $y^{\prime}$ are then parallel to these basis vectors. 

The parametrization has nine free parameters that need to be fitted. Those are the location parameters $X_+,X_-,Y_+,Y_-$, the width parameters $\sigma_+,\sigma_-$, the offset parameter $O$ and the two scaling parameters $A_+$ and $A_-$, which are positive and it holds $A_+ > A_-$. This means that the parameterization is made up of two Gaussians, which are shifted with respect to each other and subtracted from each other. 

This parameterization describes data already transformed into the shower plane. It therefore indirectly depends on a reconstruction of the direction of the shower ($\phi,\theta)$, which is needed for the transformation. This would add two parameters to the fit. However, as the arrival direction is usually measured through signal arrival times, the parameters will be available independently. 

\section{Fit quality and modification of the parameterization}
\label{sec:dis}
Function (\ref{eq:ini}) is fitted without any further restrictions to every individual simulated shower, using a standard Levenberg-Marquardt least-squares algorithm. In oder to identify suitable starting values, first one single two-dimensional Gaussian function is fitted.  This will be especially necessary if the core position (always (0,0) for simulations) is not well known, as it is typically the case for measured showers. 

The CoREAS simulations suffer from artificial signal power at large distances to the shower axis, introduced by the thinning of the simulated air showers: Particle thinning approximates several particles with an extended spatial distribution by a single particle with an appropriate weight factor. The localization of the corresponding radio source in one point leads to artificial coherence, which in turn leads to an overestimation of the radiated power at high frequencies. At large lateral distances the frequencies affected by this \emph{thinning noise} become as low as those measured by LOFAR. For an unthinned simulation, only a very low signal due to the incoherent addition of the emission from individual particles would be expected. 

Due to this thinning noise, the simulated signal power does therefore not reach zero at larger distances to the shower axis. The offset parameter $O$ is introduced to compensate this effect. It is however an additional parameter of the fit, which can induce local minima. It can be left out, at the cost of a decreased fitting quality at the outer edges of the grid. Initial tests have shown that the effect of local minima is more detrimental for the fit quality than the decreasing fit quality. It is therefore excluded from the fit, for the rest of the analysis. It might however be necessary to reintroduce this parameter for measured data, depending on the noise situation and the required signal-to-noise ratio.  

Additionally, it was found that the $Y_-$ parameter is almost constant ($Y_- < \unit[1]{m}$) for all fits and it is therefore also not needed. The full set of simulations is consequently fitted again, without the parameters $O$ and $Y_-$. The results of these fits are discussed in the following sections. 

\begin{figure}
\includegraphics[width=0.49\textwidth]{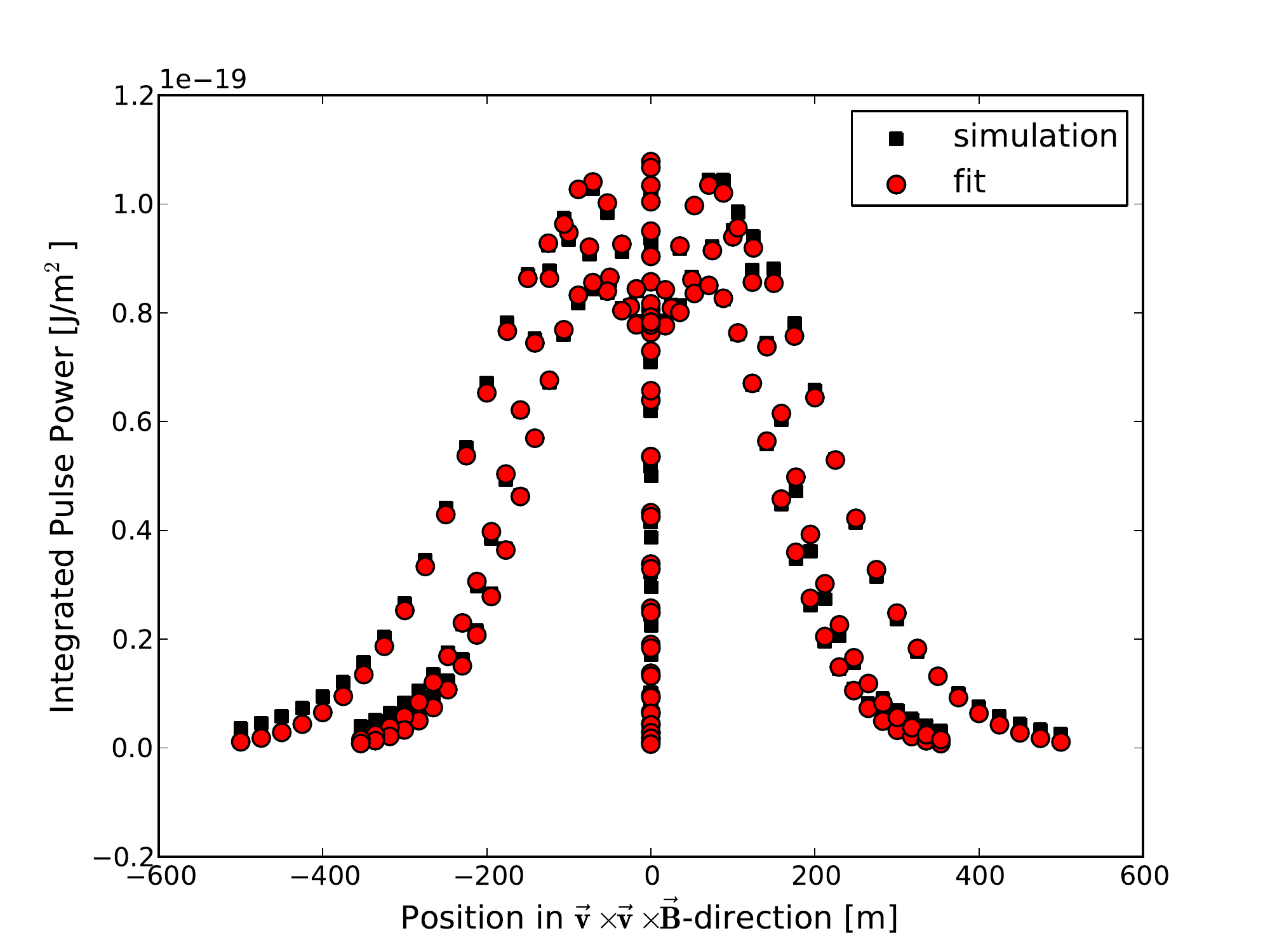}
\includegraphics[width=0.49\textwidth]{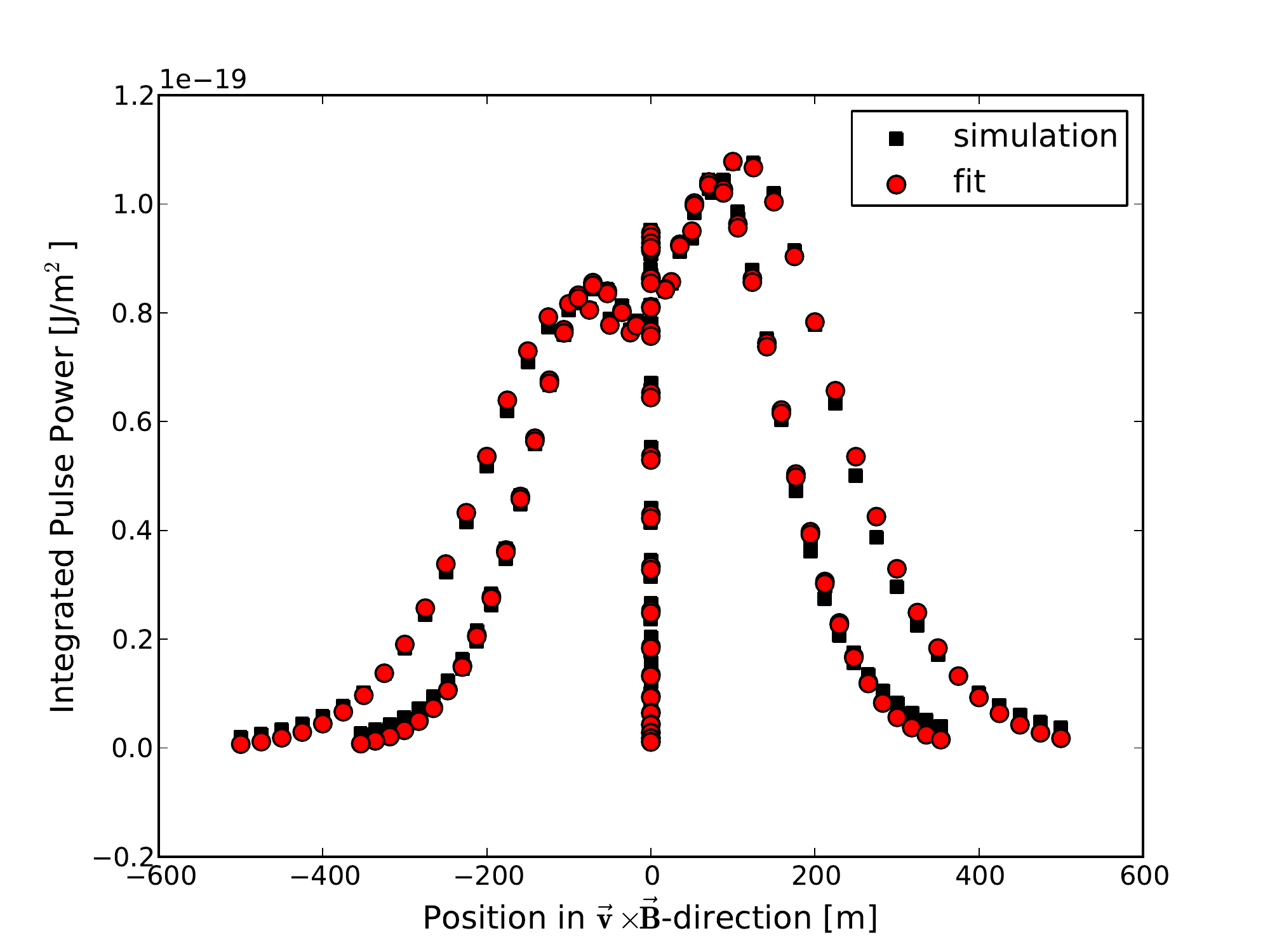}
\caption{Detailed result of the fit of a single simulated shower. In both figures the original simulation for every simulated antenna position (see figure \ref{fig:grid}) is depicted as a black square and the value of the fit at this antenna position is indicated by a red circle. The results are shown with respect to two perpendicular axes in the shower plane, thereby respectively ignoring the coordinates in the other axes. This means that the accumulation of points at 0 are those points, which lie along the perpendicular axis in the original star pattern. As the emission pattern is (almost) symmetric with respect to the $\vec{v}\times\vec{B}$-axis, the two arms of the pattern that lie at $45^{\circ}$ are only distinguishable in the left figure. This figure illustrates the agreement between the simulation and the fit to the simulation.}
\label{fig:fitdata}
\end{figure}

An example of a successful fit is shown in figure \ref{fig:fitdata}. Both, the fit and the simulated data are shown and represented as circles and squares, respectively. For better visibility cuts along the $x^{\prime}$-axis ($\vec{v}\times\vec{B}$) and the $y^{\prime}$-axis are shown, which illustrate in which coordinates the asymmetry is present. The induced electric field from the geomagnetic effect is polarized in the $\vec{v}\times\vec{B}$-direction in the same way for all antennas. For the charge excess the direction is different for all antennas, namely radially pointing towards the shower axis. Thus, there will be constructive interference for positive values of $x^{\prime}$ and destructive interference for negative values, which is visible in the cut along the  $\vec{v}\times\vec{B}$-axis ($y^{\prime}=0$). The figure shows a good agreement between simulated data and the fit. 

\begin{figure}
\includegraphics[width=0.49\textwidth]{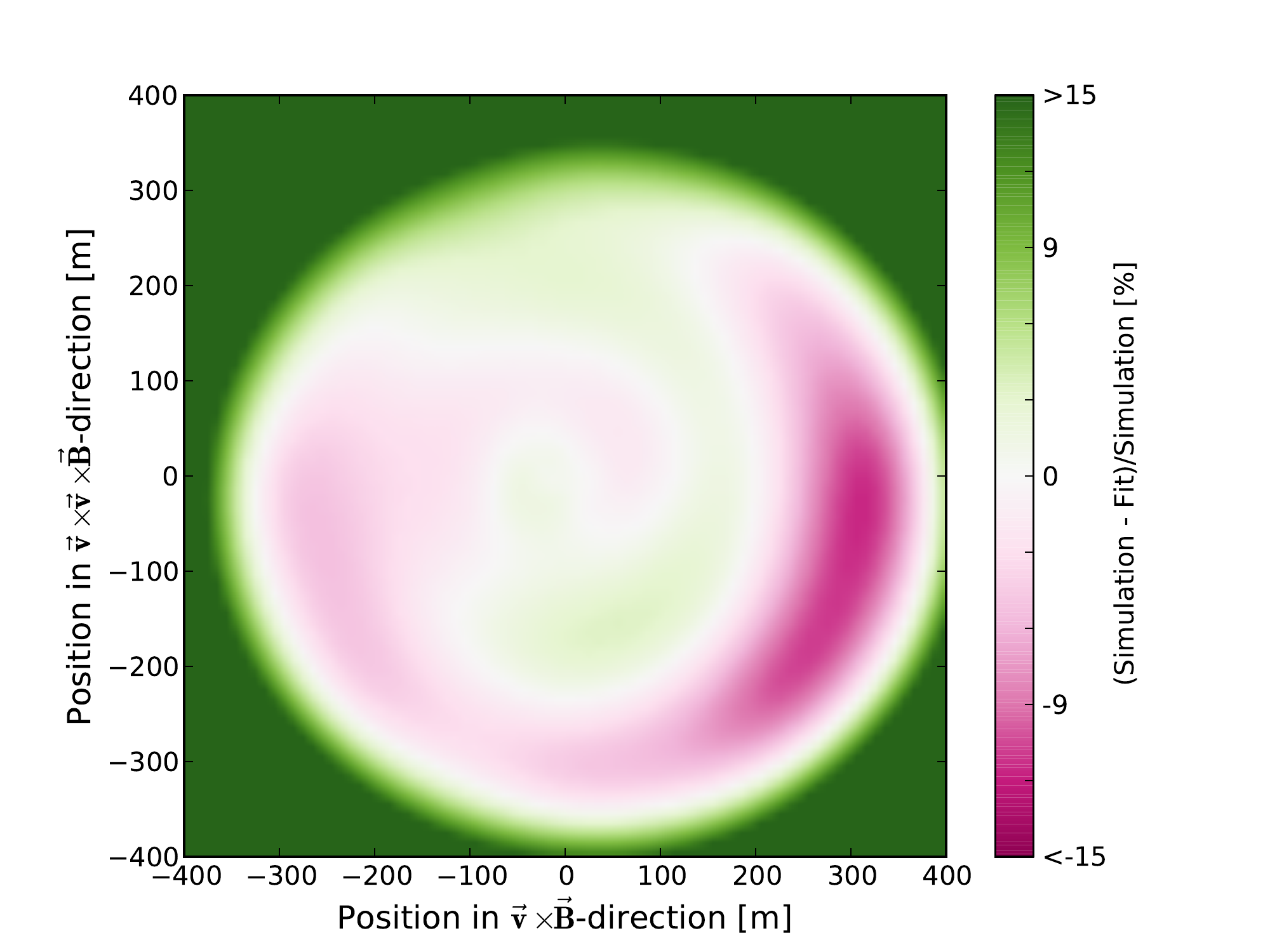}
\includegraphics[width=0.49\textwidth]{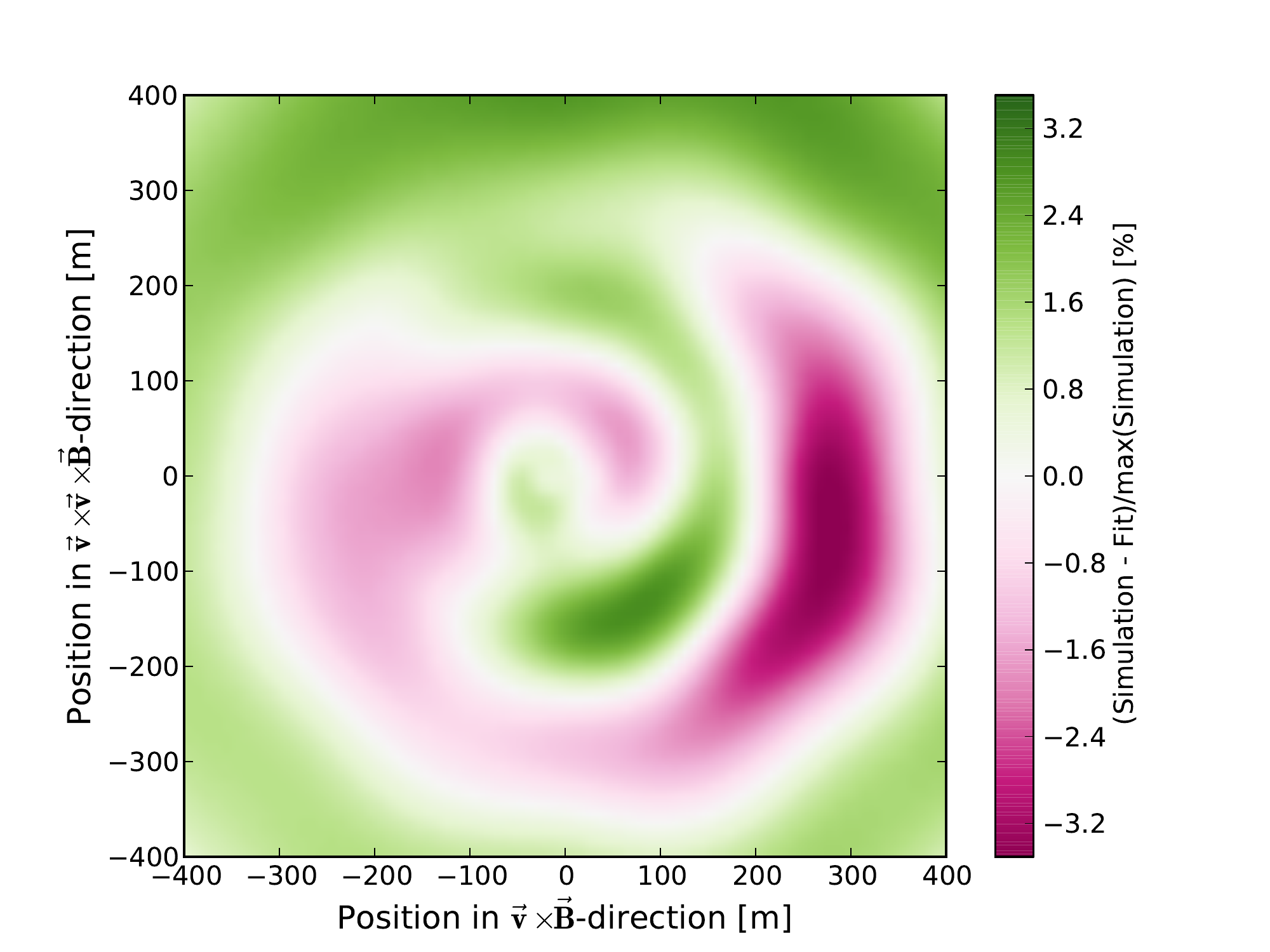}
\caption{Illustration of the residuals of the fit shown in figure \ref{fig:fitdata}. Left: Relative differences between the integrated pulse power at each antenna position as obtained from simulations and given by the fit, normalized to the value of the simulation at every position. Right: Relative differences between the integrated pulse power at each antenna position as obtained from simulations and given by the fit, normalized to the maximum integrated pulse power of the shower simulation. Features that can be interpreted as straight edges are caused by the interpolation for the plot. }
\label{fig:fitquality_single}
\end{figure}

\begin{figure}
\begin{center}
\includegraphics[width=0.6\textwidth]{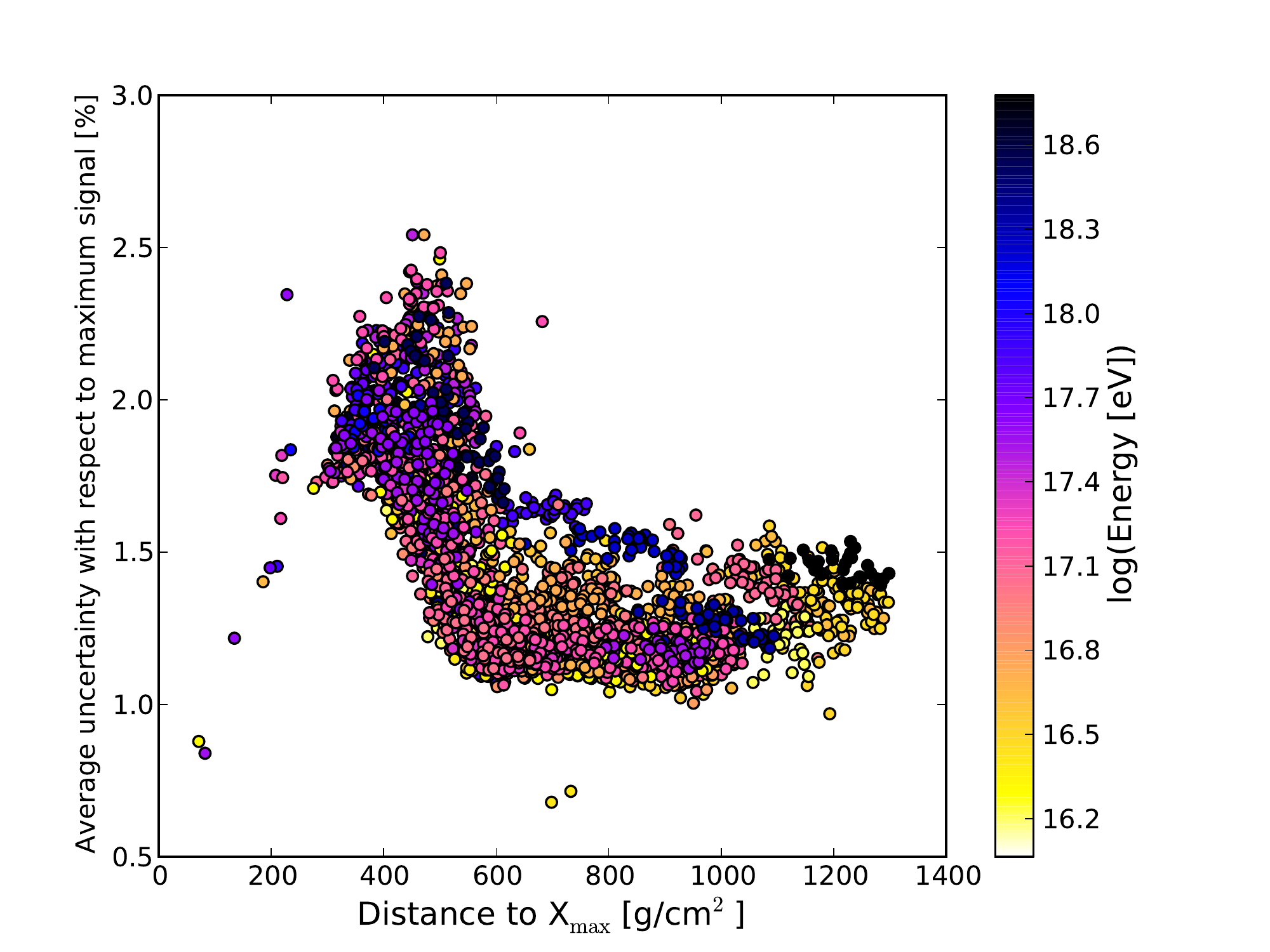}
\caption{The fit quality as a function of distance to the shower maximum (see equation (\ref{eq:d_xmax})). The uncertainty is calculated as average per simulated antenna of the absolute difference between fit and simulation. This average is re-weighted with the maximum signal in the simulation to ensure a comparability between events. The figure shows a decrease in fit quality with decreasing distance to $X_{\mathrm{max}}$. The energy of the simulated shower is encoded in color, there is no (visible) correlation with energy. }
\label{fig:fitquality}
\end{center}
\end{figure}

In order to assess the quality of the fit, the relative uncertainty is calculated. As there are no measurement uncertainties on the simulated showers, the absolute residuals are not directly comparable between events. This is especially true, given the fact that the simulated events span three orders of magnitude in energy, which delivers pulse powers that span six orders of magnitude. Therefore, the relative difference between original simulation and fit is calculated, as it is shown in figure \ref{fig:fitquality_single}. The difference between fit and simulation, normalized to the individual simulated pulse powers at every position, is shown on the left. At regions with lower signal this gets rather large as a small value is divided by an even small value. These are however the less relevant parts of the shower as they contain low (possibly experimentally not measurable) signals. In order to make the relevant part better visible the difference normalized to the maximum simulated pulse power is shown on the right. 

Those regions of the fit that show the largest deviations, are those that lie at the outer fall-off (in figures \ref{fig:fitdata} and \ref{fig:fitquality_single} at around $\unit[250]{m}$). This could be explained by the fact that the fall-off is expected to be exponential, but possibly with a different exponent. In order to obtain the observed turn-over an even exponent (2 for a Gaussian) is needed, limiting the choice of the precise slope, which results in a deviation at the fall-off. This deviation is however rather small with respect  to the other well fitted features. 

To make the uncertainties comparable, the average deviation is calculated per simulated event. This is done by determining the deviation per simulated data point with respect to the maximum signal in the shower and then averaging this deviation for the respective simulated shower. The result is shown in figure \ref{fig:fitquality}. The fit quality is not the same for all types of showers. The figure shows that the fit quality is a function of the distance to the shower maximum. The distance to the shower maximum is for this purpose defined as:

\begin{equation}
\unit[D(X_\mathrm{max})]{[g/cm^2]} = \mathrm{X_{\mathrm{atm}} [g/cm^2]} /\cos(\theta) - X_{\mathrm{max}} \mathrm{[g/cm^2]}.
\label{eq:d_xmax}
\end{equation}
Here, the column density of the distance through which the shower travels from the shower maximum is calculated. $X_{\mathrm{atm}}$ is the vertical integrated column density of the whole atmosphere. Radio emission is essentially sensitive to the geometric distance from the ground to the shower maximum (e.g. in km). For simulations with a known atmosphere, both are equivalent. 

The dependency can be explained by the effect that propagation in the atmosphere has on the radio signal. As radio emission suffers from almost no attenuation in the atmosphere, the overall detectable power stays the same with increasing distance to $X_{\mathrm{max}}$. From geometrical considerations it follows that, given a certain opening angle of the emission, the detectable power will be distributed on larger area on the ground for larger distances. If the signals are still above the detection threshold (energy threshold), that will make those larger events experimentally easier to resolve. For simulations, this means that smaller events are likely to be dominated by antennas with artificial thinning noise. This reduces the fit quality, as the Gaussian goes to zero, while the simulations do not. This could be overcome by applying a (subjective) signal-to-noise cut or reintroducing the offset parameter $O$, which induces other difficulties. Further consequences of this effect will be discussed in more detail in the following sections. 

\section{Physical interpretation of the fit parameters}
\label{sec:phys}
The fit parameters can be related to physical parameters of the shower. Each fit parameter will be discussed with respect to its primary and secondary dependencies. Correlations between parameters are likely, as the parameterization is based on the shape of the distribution rather than on possibly separable contributions to the emission. 
\subsection{Primary dependencies}
\begin{figure}
\includegraphics[width=0.49\textwidth]{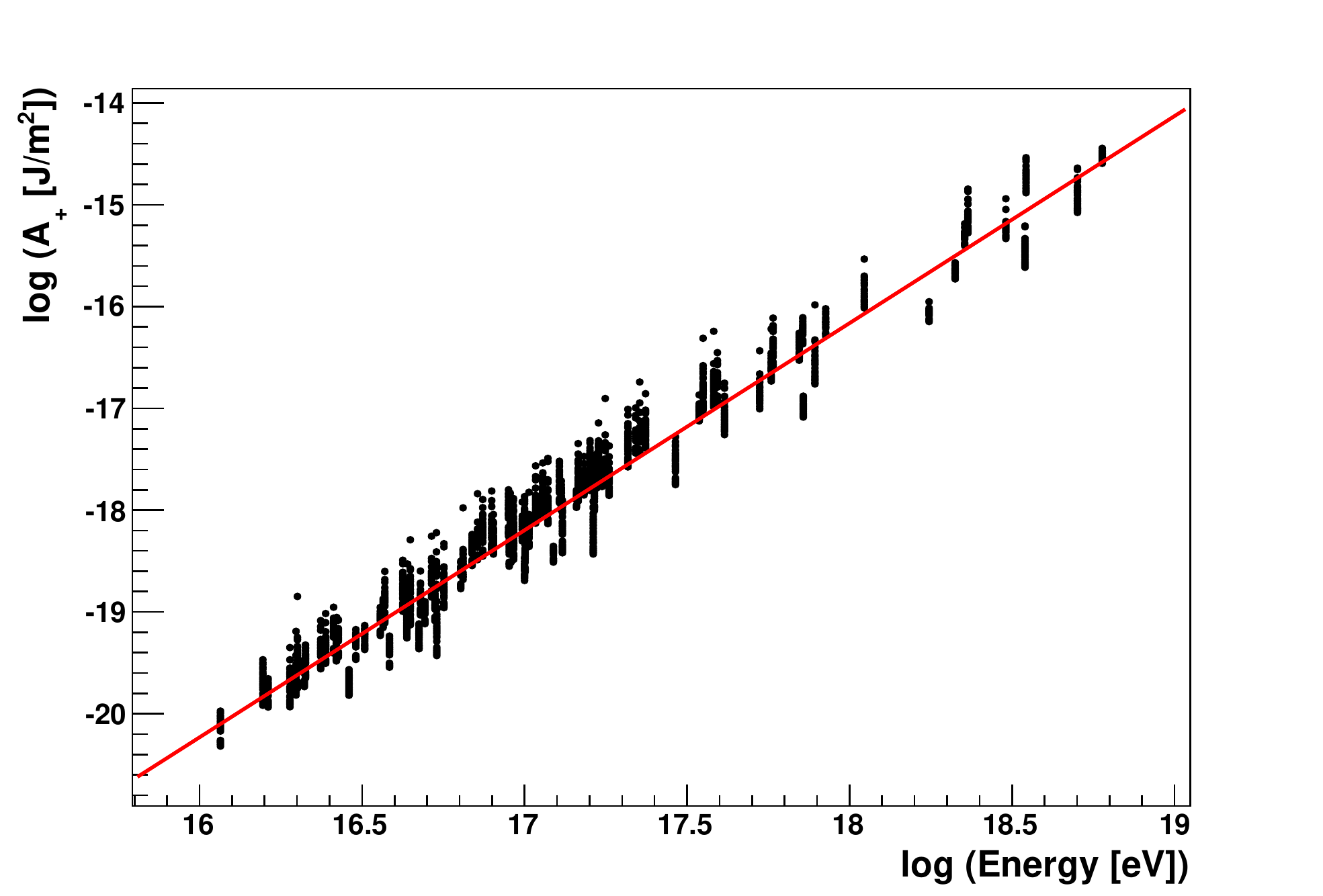}
\includegraphics[width=0.49\textwidth]{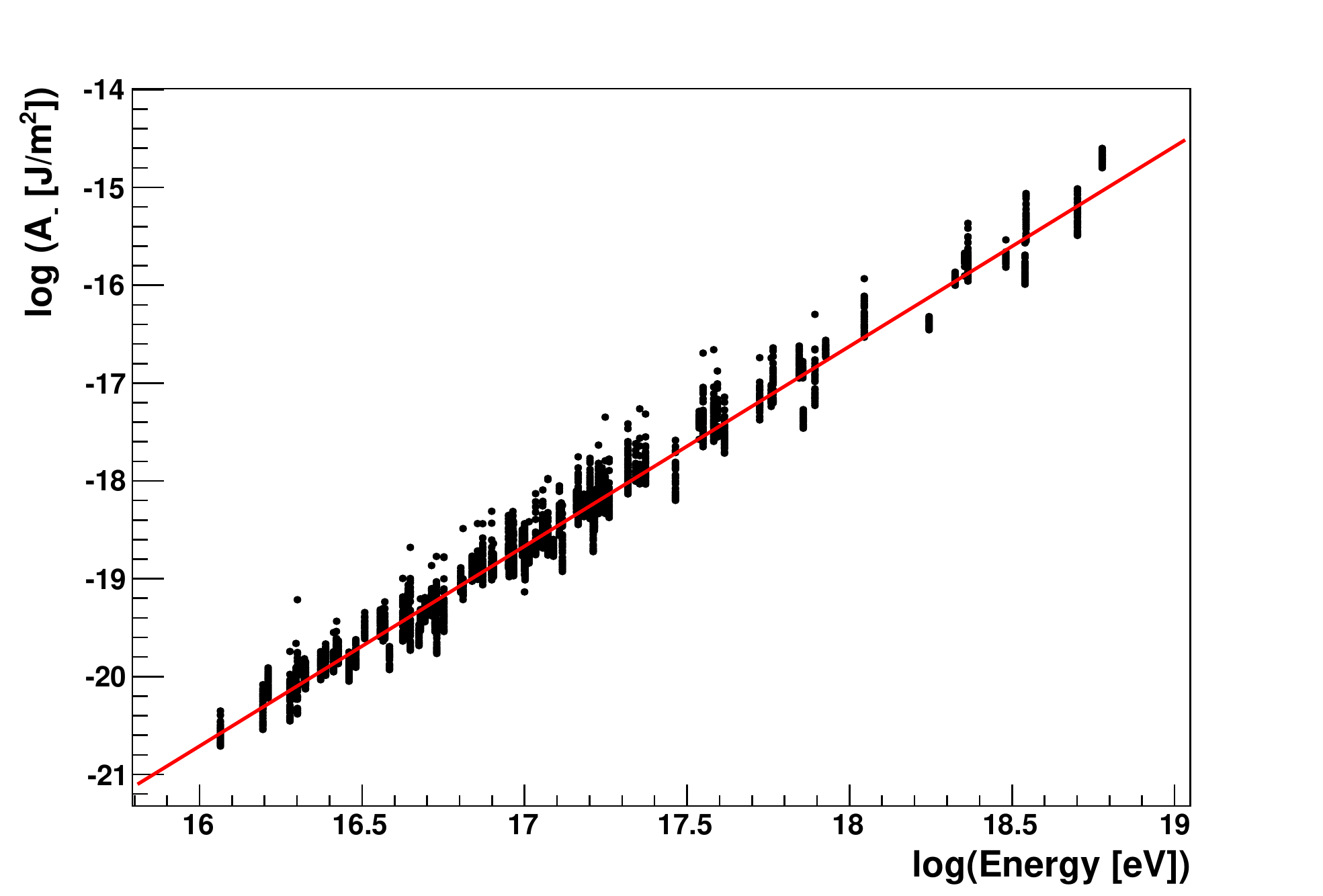}
\caption{Distribution of fitted amplitude parameters $A_+$ and $A_-$. A clear correlation with the energy of the simulated showers is shown. The red lines indicate linear fits with slopes $2.034 \pm 0.005$ and $2.042\pm0.004$, respectively.}
\label{fig:ampl}
\end{figure}

The two amplitudes $A_+$ and $A_-$ show a clear correlation with energy of the shower as shown in figure \ref{fig:ampl}. The correlation is in fact quadratic in energy, meaning that $A_{\pm} \propto E^2$, which is characteristic for coherent radiation. It was predicted  \cite{Allan1971, Huege2003} and measured \cite{Acounis2012, Revenu2013ARENA,Palmieri2013} in several studies that the amplitude of the induced electric field should indeed be proportional to the energy, from which follows a quadratic dependence for the correlation with power.  

\begin{figure}
\includegraphics[width=0.49\textwidth]{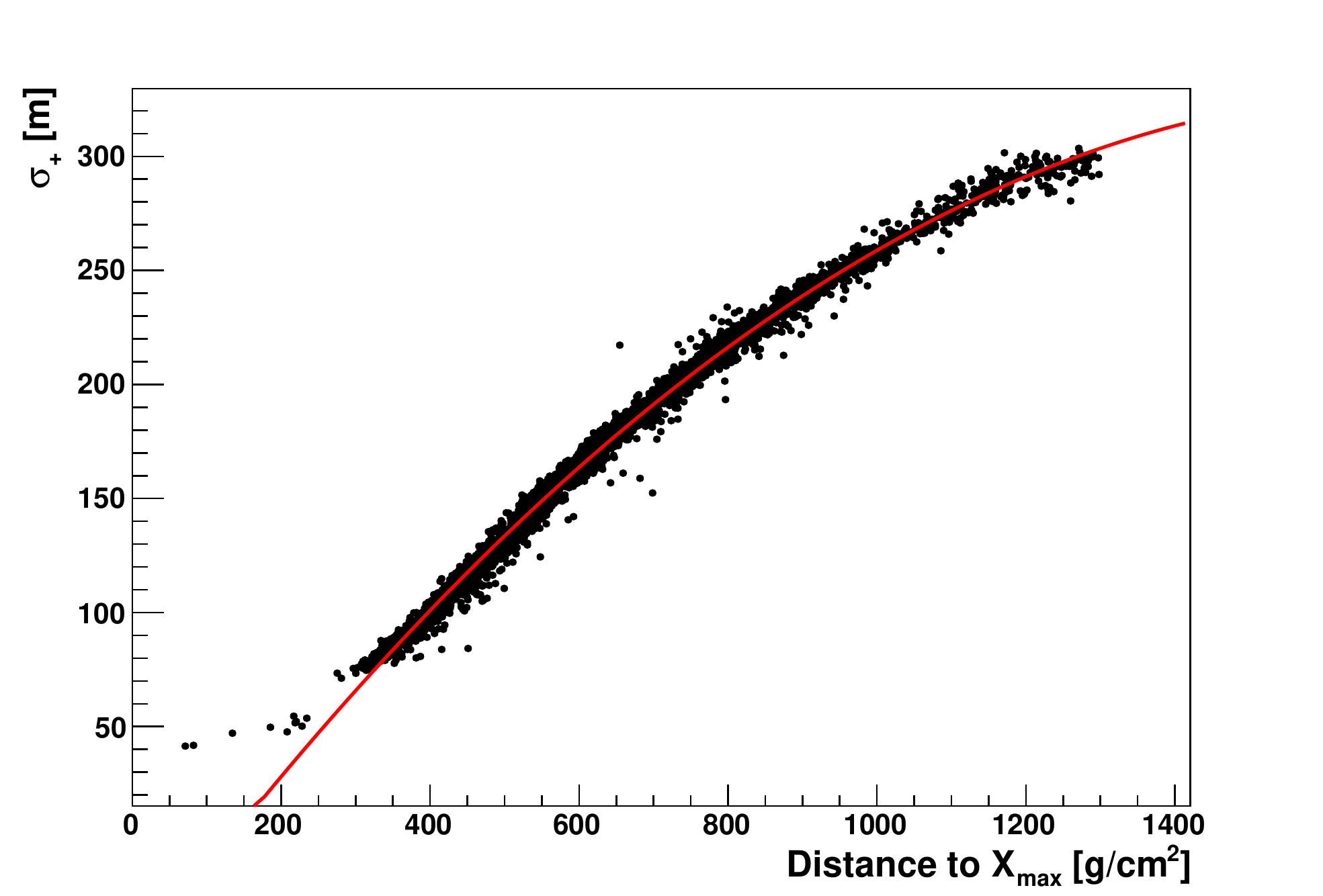}
\includegraphics[width=0.49\textwidth]{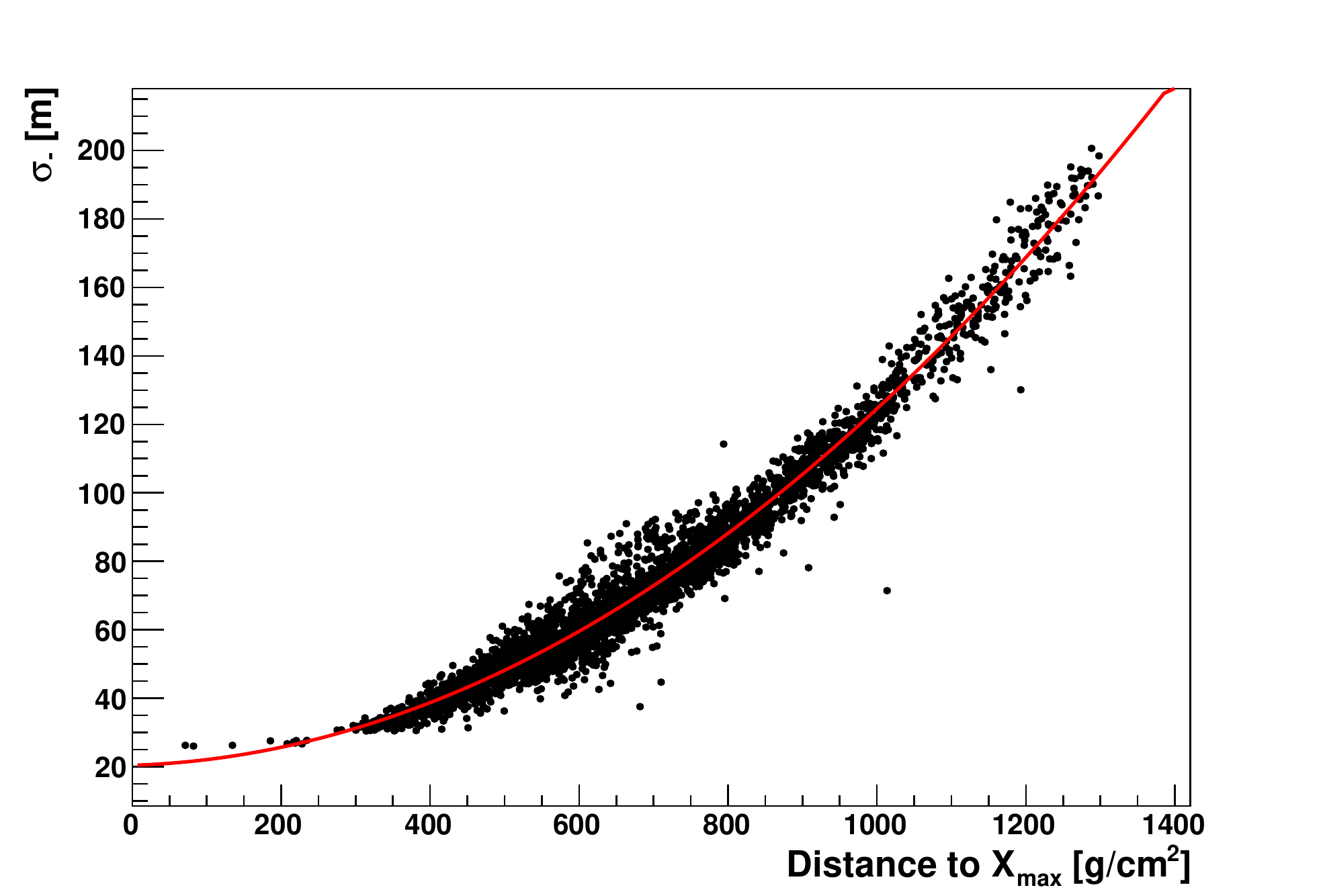}
\caption{Distribution of fitted width-parameters $\sigma_+$ and $\sigma_-$. Both parameters clearly correlate with the distance to the shower maximum. The red line indicates a second order polynomial fit. The parameters are given in the appendix.}
\label{fig:sigma}
\end{figure}

The widths of the distributions $\sigma_+$ and $\sigma_+$ show a correlation with the distance to the shower maximum, as shown in figure \ref{fig:sigma}. As already discussed before, a dependence of $\sigma_{\pm}$ on the distance to the shower maximum is expected as the signal distribution on the ground becomes wider with the propagation distance of the shower. Both parameters show a different behavior with distance to $X_{\mathrm{max}}$, however both can be described by a second order polynomial\footnote{$\sigma_+$ can also be described relatively well by a $\sqrt{x}$-function. The correlation of $\sigma_+$ and $\sigma_-$ can be described by an exponential.}.

\begin{figure}
\includegraphics[width=0.49\textwidth]{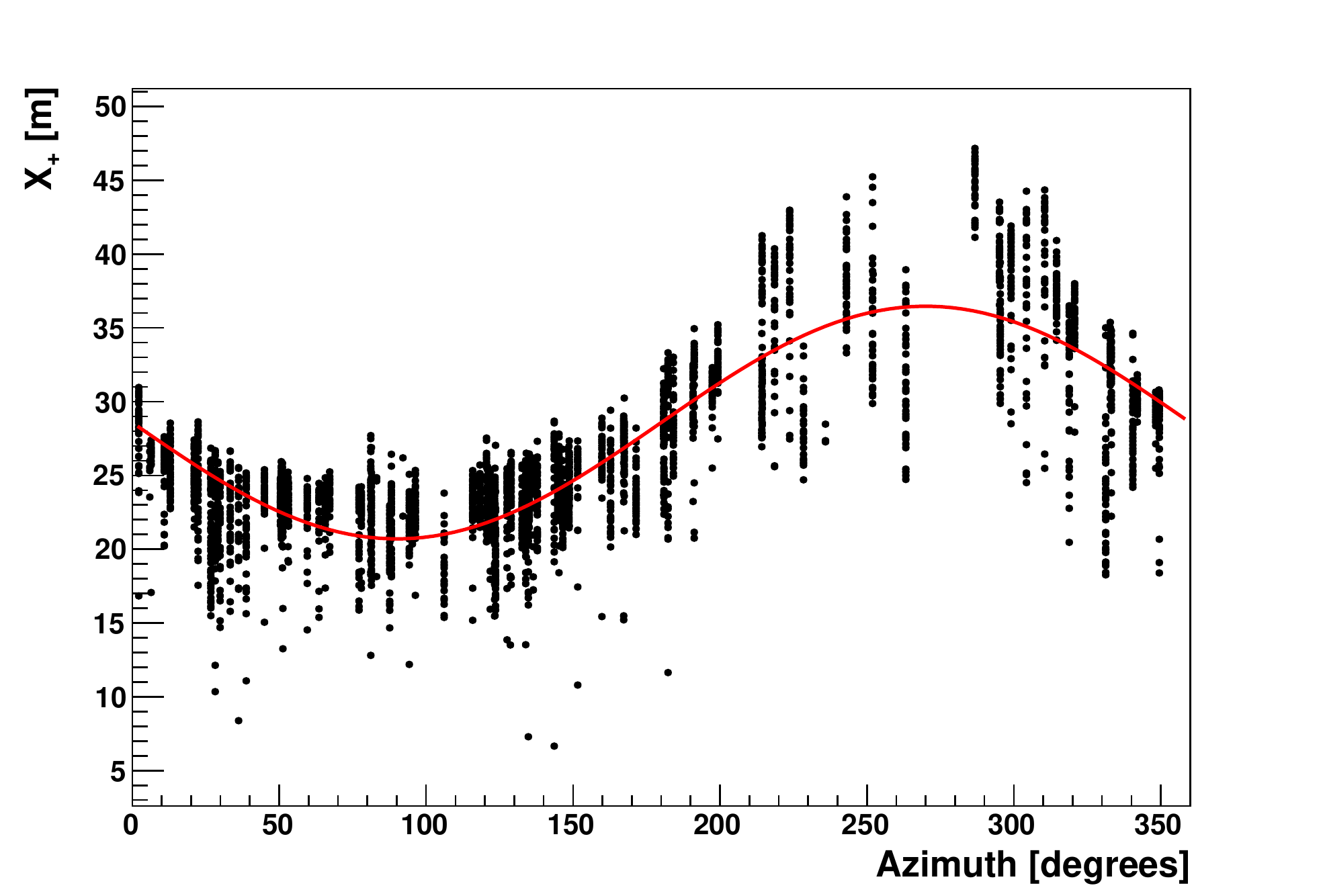}
\includegraphics[width=0.49\textwidth]{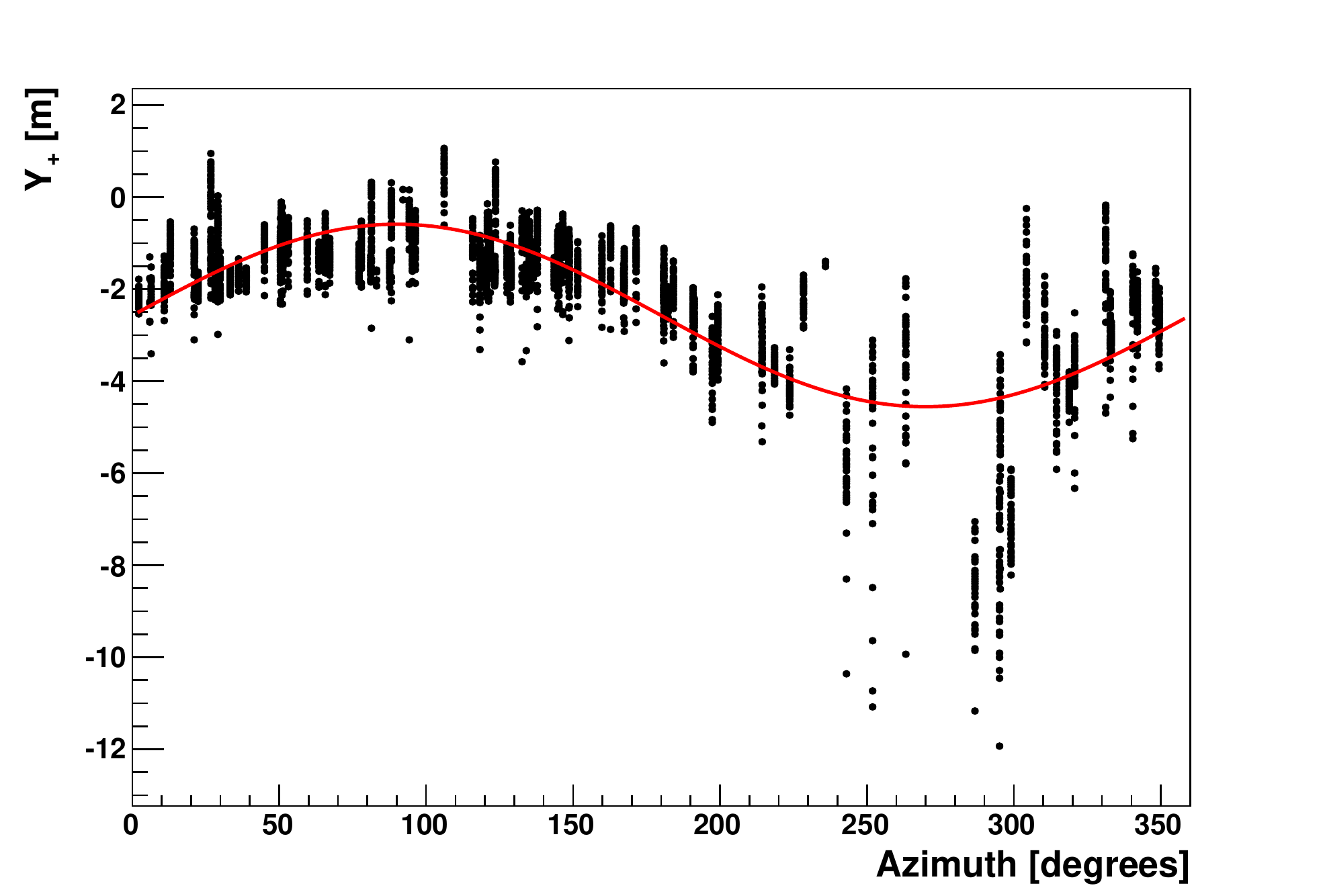}
\caption{Distribution of the fitted shift parameters $X_+$ and $Y_+$. Left: The shift of the positive Gaussian $X_+$ in the direction of the $\vec{v}\times\vec{B}$-axis shows a dependence on azimuth angle.  Right:  The shift of the positive Gaussian $Y_+$ in the direction of the $\vec{v}\times \vec{v}\times\vec{B}$-axis also shows an, although different, dependence on azimuth angle. Sine functions are fitted to both parameters.}
\label{fig:largeGauss}
\end{figure}

The additional parameters show less clear, nonetheless interesting dependencies. The left side of figure \ref{fig:largeGauss} shows the main dependence of the shift of the positive Gaussian with respect to the shower core. The shift depends on the sine of the azimuth of the arrival direction of the shower, which is measured in coordinates on the ground. This is most likely related to the interplay between the charge excess and the geomagnetic signal contribution. At azimuth angles of $90^{\circ}$ (North) the arriving air showers are more perpendicular to the magnetic field, meaning that the geomagnetic contribution is stronger than at $270^{\circ}$ (South), where the showers are more parallel to the local magnetic field. Given that the contribution of the charge excess is equally strong for all arrival directions, the ratio of the two processes changes with azimuth angle. The lateral fall-off of the two contributions is expected to be different \cite{Vries2010}. The charge excess falls off flatter than the geomagnetic contribution. The addition of the two effects is therefore different for different observer positions with respect to the shower core. In the observer direction in which the electric fields of the two contributions are parallel (positive $\vec{v}\times\vec{B}$), this means that once the ratio of the two contribution shifts towards more charge excess, the maximum signal will move further out with respect to the core. This is what the fit represents in the $X_+$ parameter. 

The dependencies of $Y_+$  on the arrival direction as shown on the right in figure \ref{fig:largeGauss} are significantly smaller. The changes as function of azimuth angle  are the largest for events arriving from the direction of the magnetic field ($270^{\circ}$). Additionally, the parameter shows a weak dependence on the zenith angle of the arrival direction: the scatter around the mean value increases with zenith angle. One could speculate about the physical explanation of this behavior as function of the shower development. As, however, shifts of less than $\unit[5]{m}$ are usually experimentally not resolvable in air showers, the dependencies are not relevant for the applicability of the parameterization. 

\begin{figure}
\begin{center}
\includegraphics[width=0.49\textwidth]{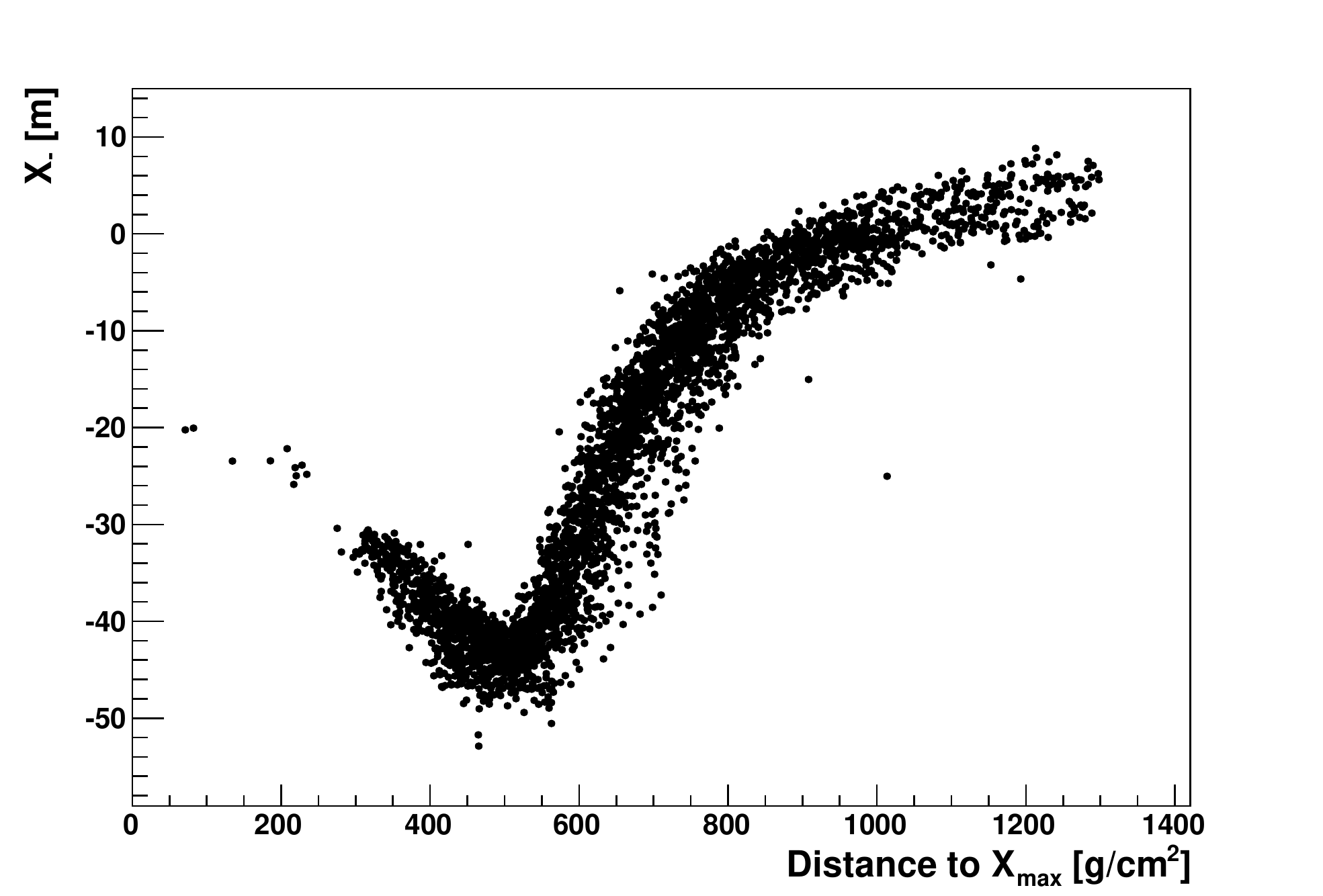}
\caption{Distribution of the fitted shift parameter $X_-$. This parameter shows a dependence on the distance to shower maximum. }
\label{fig:smallshift}
\end{center}
\end{figure}

The most difficult dependence to explain is the one of $X_-$ as it is shown in figure \ref{fig:smallshift}. For large values of distance to the shower maximum, it shows a rise similar to the one observed in $\sigma_+$. This  however changes for values smaller than $\unit[500]{g/cm^2}$. There are two possible explanations. Either this point coincides with the point at which the quality of the fit decreases (see figure \ref{fig:fitquality}). As discussed before, the decreased quality might be correlated with the number of antennas with a significant signal. It is easy to conclude that once the area with significant signal becomes smaller, the importance of the negative Gaussian will decrease. If the structure of the peak is no longer well resolved, the values for the position of the negative Gaussian will be less  well defined. Alternatively, we could be observing a change in the dominance of processes. The negative Gaussian is needed to capture the asymmetry of the signal due to charge excess and geomagnetic effect, as well as the enhancement induced by the relativistic time compression (here, ring like structure at larger zenith angles.). If for events that penetrate deep into the atmosphere the enhancement due to time compression becomes less important, the $X_-$parameter might also change behavior. It should be noted that it is possible, however very unlikely, that this behavior is artificial and introduced by the choice of fitting procedure. Testing a different algorithm has lead to comparable results.

\subsection{Secondary dependencies}
If one corrects the fit parameters by the aforementioned relations, one can observe secondary dependencies on air shower parameters.  

\begin{figure}
\includegraphics[width=0.49\textwidth]{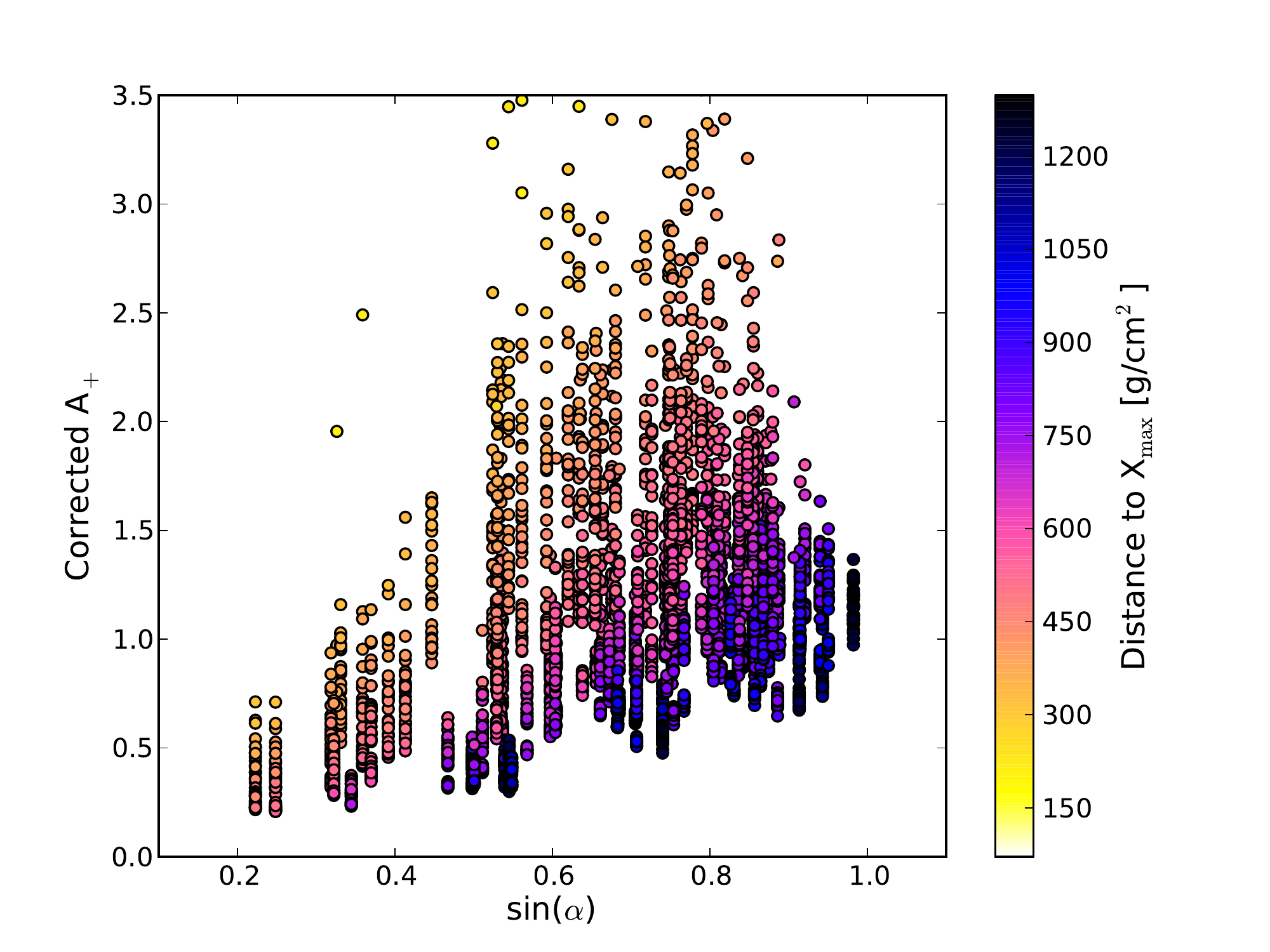}
\includegraphics[width=0.49\textwidth]{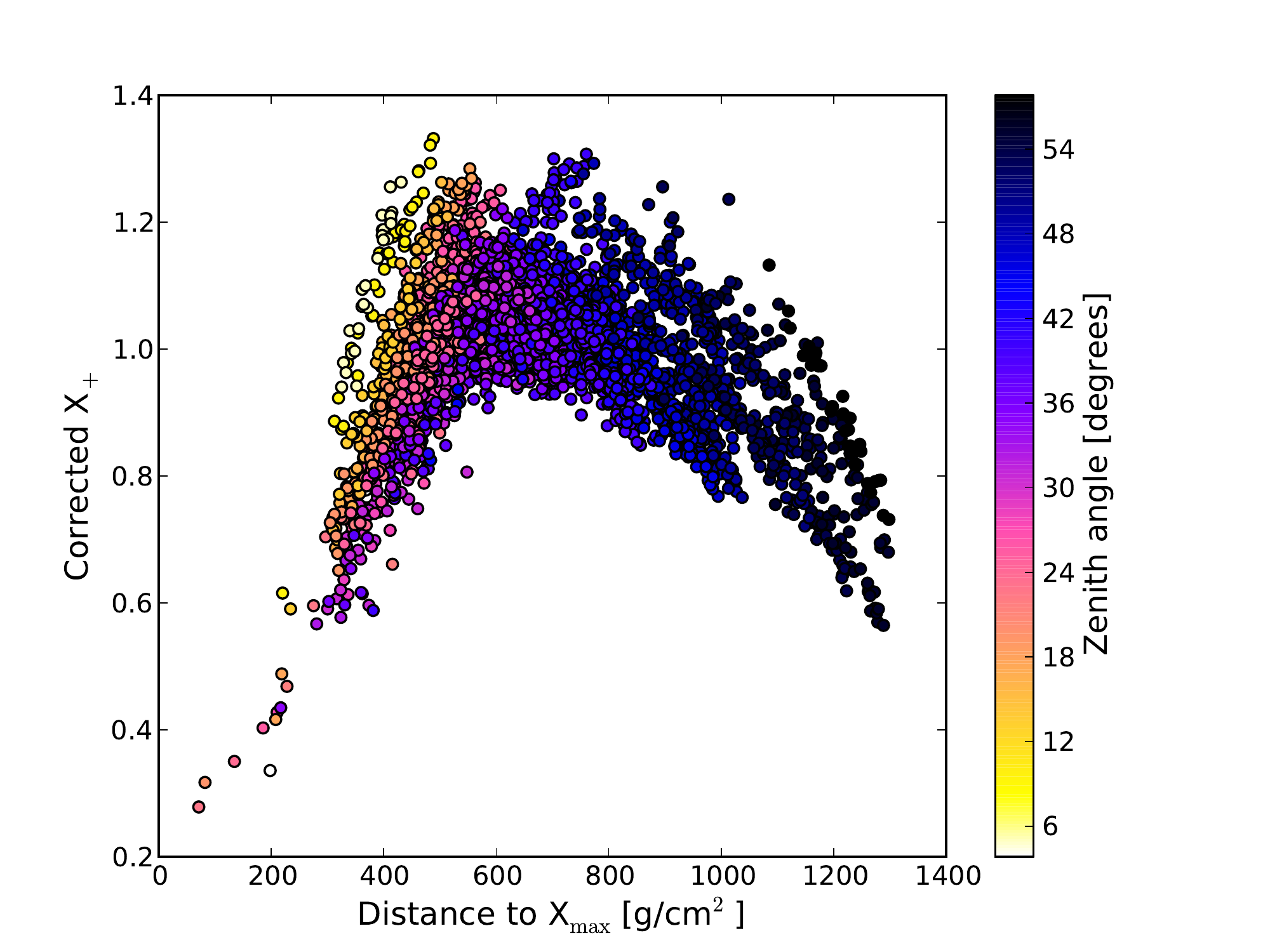}
\caption{Distribution of secondary dependencies on shower parameters. Left: The parameter $A_+$ is corrected for the dependence on energy. The remaining scatter is a function of the angle between the shower axis and the direction of the magnetic field $\alpha$ and the distance to the shower maximum,  $X_{\mathrm{max}}$. The dependence on $\sin(\alpha)$ also varies with different distances to $X_{\mathrm{max}}$. Right: The parameter $X_+$ is corrected for the dependence on the arrival direction. The remaining scatter shows a geometric dependence on $X_{\mathrm{max}}$.}
\label{fig:ampl_res}
\end{figure}

The remaining scatter of $A_{\pm}$ shows a dependence on the angle $\alpha$ between the shower axis and the local magnetic field, as already suggested in \cite{Allan1971}. This is visualized in figure \ref{fig:ampl_res}. However, what is also shown is that the distance to the shower maximum has a large influence, which is not taken into account in earlier parameterizations. This dependence is already visible in figure \ref{fig:ampl} by the vertical groups of points, indicating an air shower of the same energy and direction with different values for the shower maximum. Figure  \ref{fig:ampl_res} shows that determining the energy of the shower solely based on $A_{\pm}$  is not the approach that delivers the highest resolution. The energy resolution based on $A_{\pm}$ will however improve, if one uses the independent parameter of the angle to the magnetic field (obtained from the arrival times of the radio signals in the antennas) and an estimate of the distance to the shower maximum, as obtained from $\sigma_{\pm}$. In order to resolve the energy with a higher accuracy, a combination of parameters or the power at a certain distance might be worth pursuing \cite{HuegeUlrichEngel,Palmieri2013}.

The remaining fluctuations for $\sigma_{\pm}$ are $8\%$ effects and show no obvious secondary correlation with other air shower parameters. Detailed studies of the resolution achievable for the energy and  $X_{\mathrm{max}}$ will be discussed in a forthcoming publication and will include systematic as well as experimental uncertainties. 

\begin{figure}
\begin{center}
\includegraphics[width=0.49\textwidth]{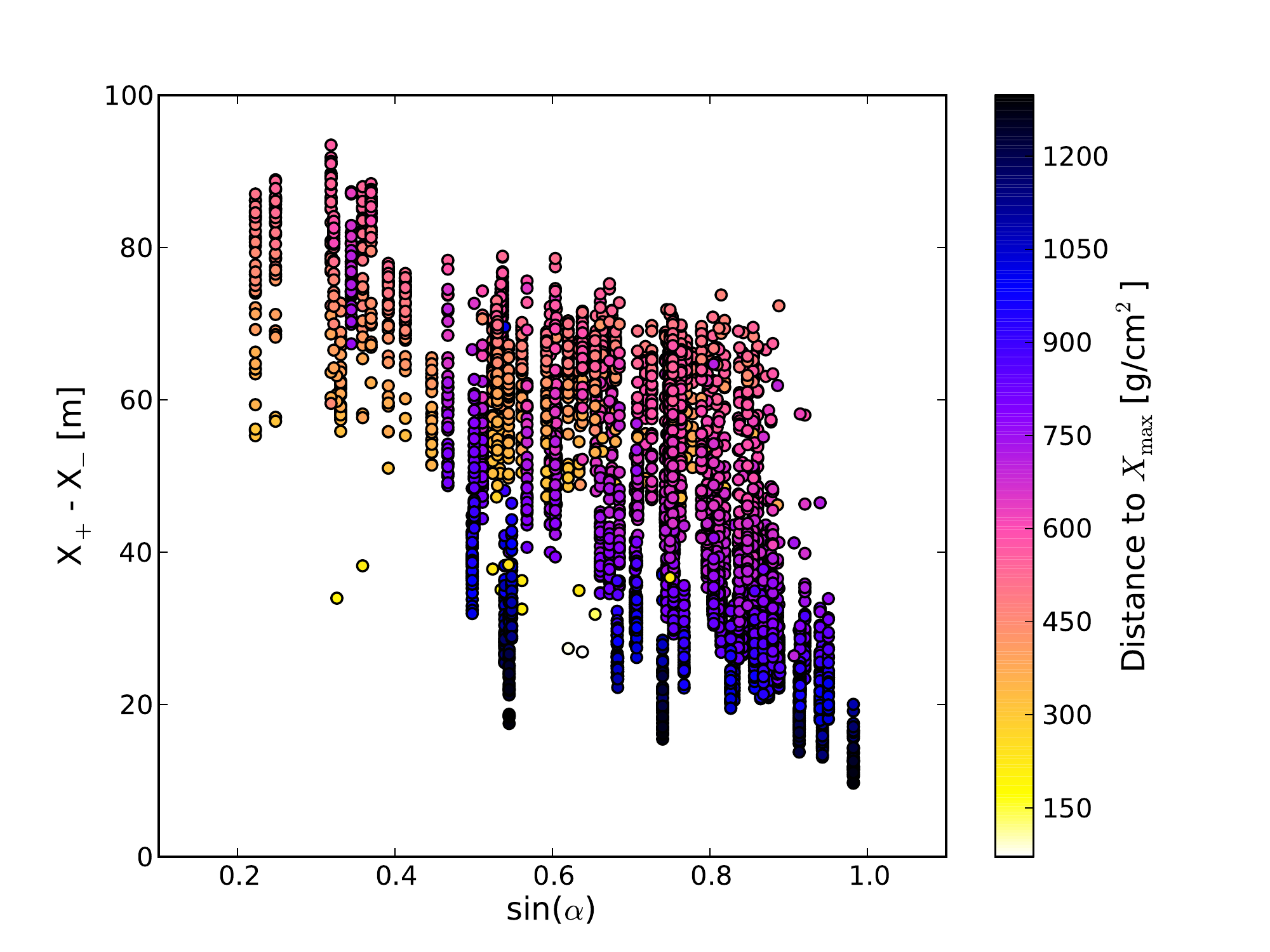}
\end{center}
\caption{Dependence of the difference of $X_+$ and $X_-$ as a function of the angle between the shower and the magnetic field. The dependence on the distance to the shower maximum is encoded in color.}
\label{fig:dep_x}
\end{figure}

The residuals from the sine fit to $X_+$ are about $15\%$ and show a slight dependence on distance to $X_{\mathrm{max}}$, as shown in figure \ref{fig:ampl_res}. It is interesting to note that they show a different behavior for different values of $X_{\mathrm{max}}$. For events with large zenith angles (horizontal showers) the correction factor is underestimated for small values of $X_{\mathrm{max}}$ (high showers) and for small zenith angles (vertical showers) the correction factor is overestimated for small values of $X_{\mathrm{max}}$. This could be explained by the fact that this fitting variable does not only represent one single mechanism but a combination, also including relativistic time compression, which correlate with the zenith angle, as it was already discussed for $X_-$. As $X_+$ and $X_-$ both describe shifts in the $\vec{v}\times\vec{B}$-axis, a correlation of the two parameters with respect to the change in dominating effect is likely. This is illustrated in figure \ref{fig:dep_x}. Here, the difference between the two parameters, i.e. the offset of the two Gaussians with respect to each other, is shown.  This offset mainly depends on the angle $\alpha$ between the shower and the magnetic field. It therefore describes the interference between the emissions from geomagnetic effect and the charge excess. At high values of $\alpha$ the geomagnetic effect dominates and the difference becomes small. Additionally, the difference depends on the distance to the shower maximum, as it is encoded in color. This dependence changes direction for increasing $\alpha$: at low $\alpha$ showers with a large distance to the shower maximum have the larger offset, at higher $\alpha$ this is reverted. This is an additional argument for the observed change in behavior. It should however be noted that the secondary dependency on distance to $X_{\mathrm{max}}$ ($15\%$ of $\unit[35]{m}$) is probably experimentally unresolvable. 

\section{Reduction of the parameterization and test on data}
\label{sec:data}
Using the findings of the previous section, the initial parameterization can be reduced in two ways.

\subsection{Direct reduction to a function of air shower properties}
One can chose to rewrite the equation (\ref{eq:ini}) in a way that it is only a function of physical shower parameters, namely the energy of the shower $E$, the arrival direction $(\theta, \phi)$, the position of the shower maximum $X_{\mathrm{max}}$ and the position of the shower axis $(X,Y)$:

\begin{equation}
P(x^{\prime},y^{\prime}) = f_1(E) \cdot \exp\left(\frac{-f_2(\phi,X,Y,x^{\prime},y^{\prime})}{f_3(\theta,X_{\mathrm{max}})}\right) - C_0 \cdot f_1(E)\cdot  \exp\left(\frac{-f_4(\theta,X_{\mathrm{max}},X,Y,x^{\prime},y^{\prime})}{f_5(\theta,X_{\mathrm{max}})}\right) 
\label{eq:full}
\end{equation}
with 
\begin{eqnarray}
f_1(E) &=& C_1 \cdot E^2,\\
f_2(\phi,X,Y,x^{\prime},y^{\prime})&=& [x^{\prime}-(X+C_2\cdot \sin(\phi)+C_3)]^2+[y^{\prime}-(Y +C_4\cdot \sin(\phi)+C_5)]^2,\label{eq:X}\\
f_3(\theta, X_{\mathrm{max}})& = & [C_6 + C_7\cdot (X_\mathrm{atm}/{\cos(\theta)}-X_{\mathrm{max}})+C_8\cdot (X_\mathrm{atm}/{\cos(\theta)}-X_{\mathrm{max}})^2]^2 ,\\
f_4(\theta,X_{\mathrm{max}},X,Y,x^{\prime},y^{\prime})&=& [x^{\prime}-(X+\sum\limits^{4}_{n=0} C_{12,n} \cdot (X_\mathrm{atm}/\cos(\theta)-X_{\mathrm{max}})^{n})]^2+(y^{\prime}-Y)^2,\label{eq:Y}\\
f_5(\theta, X_{\mathrm{max}})& = & [C_9 + C_{10}\cdot (X_\mathrm{atm}/\cos(\theta) -X_{\mathrm{max}})+C_{11}\cdot (X_\mathrm{atm}/\cos(\theta)-X_{\mathrm{max}})^2]^2
\end{eqnarray}
The constants $C_0,\ldots C_{11}$ and $C_{12,0} \ldots C_{12,4}$ have to be determined from simulation studies and can be found exemplary in the appendix (table \ref{tab:parameters}) for the LOFAR conditions, in particular the frequency range of the bandpass filter, the direction and strength of geomagnetic field, and the altitude above sea-level. It should be noted that for real measurements a changing atmosphere has to be taken into account to determine the height of shower maximum in $\unit{g/cm^2}$ and the experimental dependence is more likely to be the actual physical distance in $\unit{km}$. For simulations, which are all made using the same atmosphere, those two parameters can be translated into each other with a known relation, based on the atmospheric model used. 

\begin{figure}
\begin{center}
\includegraphics[width=0.49\textwidth]{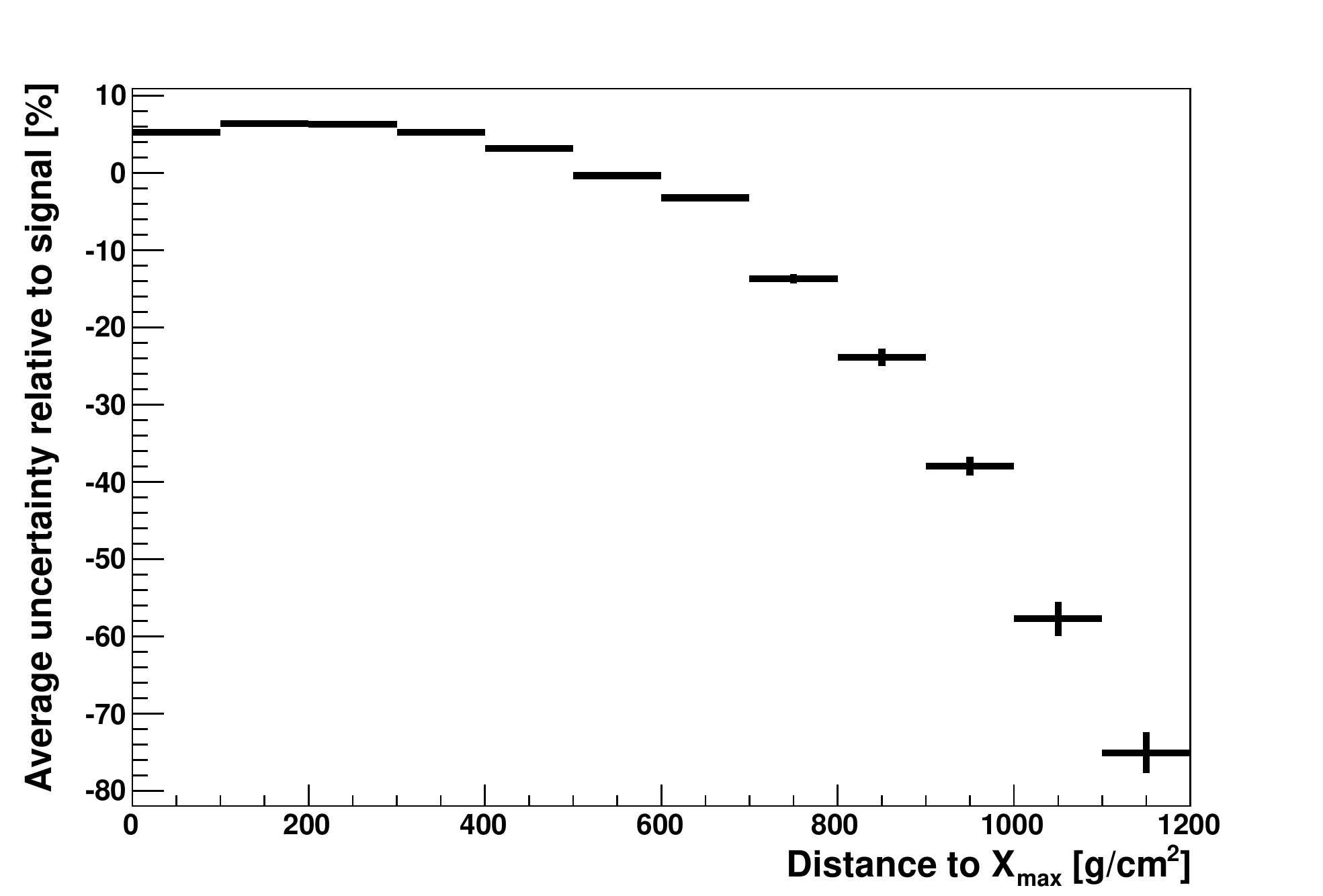}
\end{center}
\caption{Profile of the residuals between the prediction of equation (\ref{eq:full}) and the original simulation as function of distance to shower maximum. The residuals are calculated individually for all simulations and made comparable by a scaling with respect to the maximum signal. Every bin contains all showers of the simulated set with the same distance to the shower maximum. No selection on other shower parameters is applied. The quality of the prediction on average reduces for showers with larger distances to $X_{\mathrm{max}}$. The average absolute residual for all simulations is 16.9\%. }
\label{fig:reverse_fit}
\end{figure}

This function can be used to predict the radio signal from a given air shower, using a given arrival direction, energy and $X_{\mathrm{max}}$. Applying equation (\ref{eq:full}) to the same simulation set as from which it was derived, results in a measure of the quality of this description. This is shown in figure \ref{fig:reverse_fit}. The relative residuals are on average ($\pm$)16.9\% of the signal. The uncertainty is dominated by the scaling factor $A_+$. The spread in the prediction of the parameter $A_{\pm}$ has an additional strong dependence on the shower maximum, as it is illustrated in figure \ref{fig:ampl_res}. For showers with small distances to $X_{\mathrm{max}}$, $A_+$ is underestimated, for large distances it is overestimated. The prediction quality is only a function the distance to the shower maximum and thereby of zenith angle and $X_{\mathrm{max}}$ of the shower. 

A study of the individual simulations shows that the width and location parameters are very well predicted, when compared to the original fit of the simulations.  Only the estimate for the absolute scaling is lacking. The prediction for $A_+$ could be improved if one used different  $f_1(E)$ for different zenith angle regimes or bins of distance to $X_{\mathrm{max}}$, which would also allow for a specific correction for the dependence on the angle to the magnetic field.  

Due to the method it was derived with, the parameterization describes those air showers best that occur most frequently in the set of simulations: events with zenith angles between $30^{\circ}$ and $45^{\circ}$ that have an average value of $X_{\mathrm{max}}$. If the set is restricted to these parameters, the average uncertainty of the prediction reduces to less than 10\%.  This is clearly sufficient for a fast prediction of the signal distribution of the most common air showers. 

\subsection{Reduction to a stable function for data analysis}
Regarding the fit stability, equation (\ref{eq:full}) is a less optimal choice as it is prone terminate in local minima during fitting, given the multiple occurrence of the same fit parameters. Furthermore, given the  uncertainties of total amplitude calibration of experiments, as well as atmospheric models, the initial equation can be reduced in a different way, less dependent of the absolute scale of the simulation results. One can use the above mentioned relations to exploit correlations between parameters. As for example both $\sigma_+$ and $\sigma_-$ only depend on the distance to the shower maximum, they are also a function of each other. The relation is best and easiest described by an exponential dependence. This results in the following parameterization:
 
\begin{equation}
P(x^{\prime},y^{\prime}) = A_+ \cdot \exp\left(\frac{-[(x^{\prime}-X_c)^2+(y^{\prime}-Y_c)^2]}{\sigma_+^2}\right) - C_0\cdot A_+\cdot  \exp\left(\frac{-[(x^{\prime}-(X_c+x_-))^2+((y^{\prime}-Y_c))^2]}{(C_1\cdot e^{C_2\cdot \sigma_+})^2}\right)
\label{eq:fit}
\end{equation}
Here, $(X_c, Y_c)$ is the position of the large Gaussian and does not coincide with the location of the shower axis of the particle component of the shower. That axis can be inferred by using the relation in (\ref{eq:X}). Together with the known direction parameters ($\theta$, $\phi$), this reduced approach leaves five free parameters: The position $(X_c,Y_c)$, the scaling factor $A_+$, the width factor $\sigma_+$ and an offset factor $x_-$.  $C_0$, the ratio between $A_+$ and $A_-$,  is an almost constant, but non-linear function of the distance to the shower maximum. Allowing the constant $C_0$ to vary in a small range, will improve the fit quality, but will not significantly affect the resulting other parameters.

In order to minimize the number of parameters, $x_-$ could also be fixed to an average value. This would assume a typical value for $X_{\mathrm{max}}$\footnote{This is known for example in the description of particle showers, where an average shower shape parameter is chosen in the lateral distribution function, which ignores the dependence on the shower maximum}.  Especially for experiments with a large spacing between antennas and therefore small number of measurements per shower, a reduction of parameters could prove useful, if using the parameterization to determine the geometry or an energy estimation. For a model reduced in such a way, one would only need a minimum of four independent measurements.  Fixing the parameter would however come at the cost of a reduced fit quality. Giving the set of simulations used for this analysis, it can be stated that air showers with a large number of measurements cannot be fitted with a good quality with a fixed $x_-$ parameter. 

\subsection{Test on LOFAR data}

\begin{figure}
\begin{center}
\includegraphics[width=0.52\textwidth]{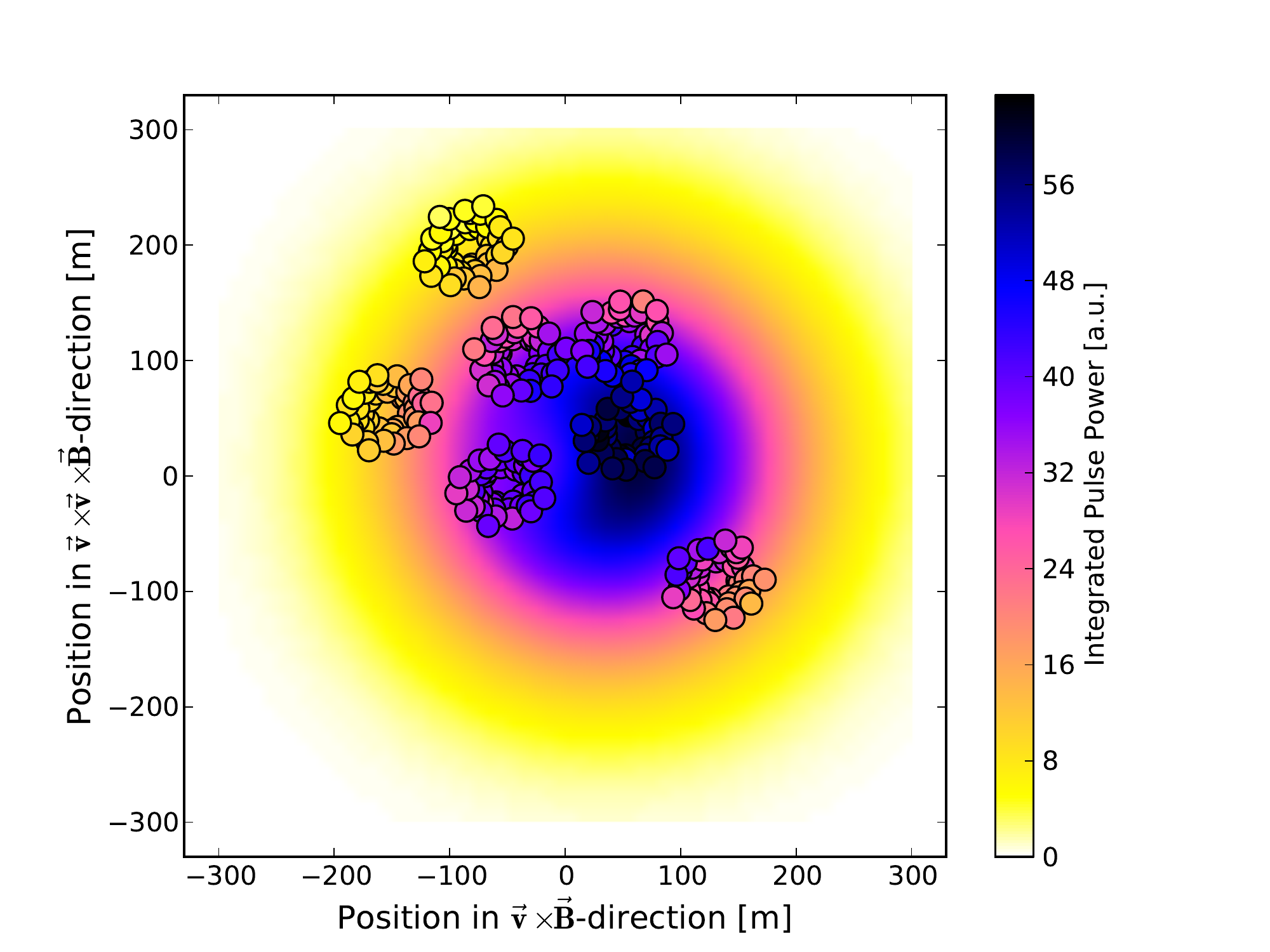}
\includegraphics[width=0.47\textwidth]{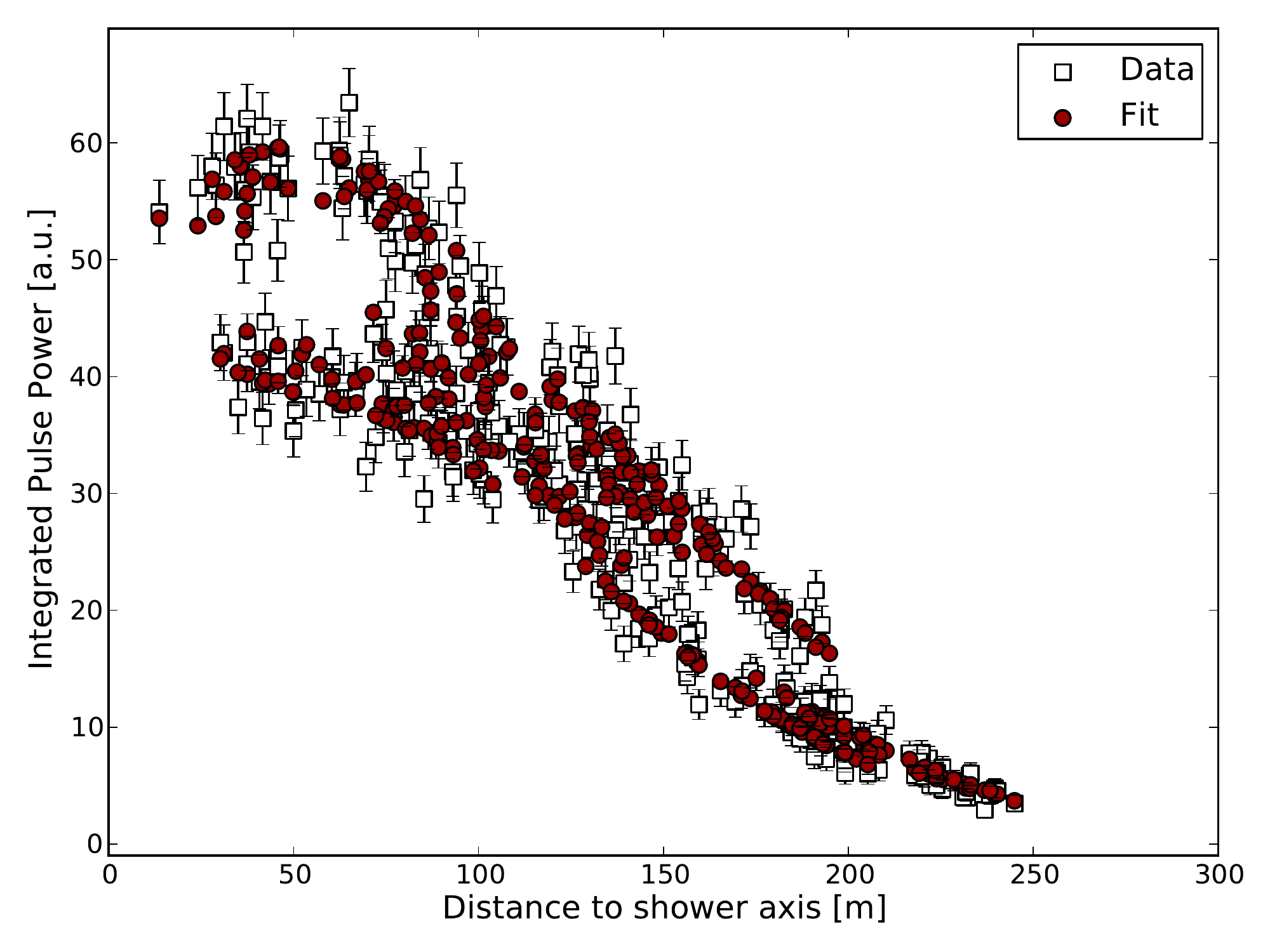}
\includegraphics[width=0.52\textwidth]{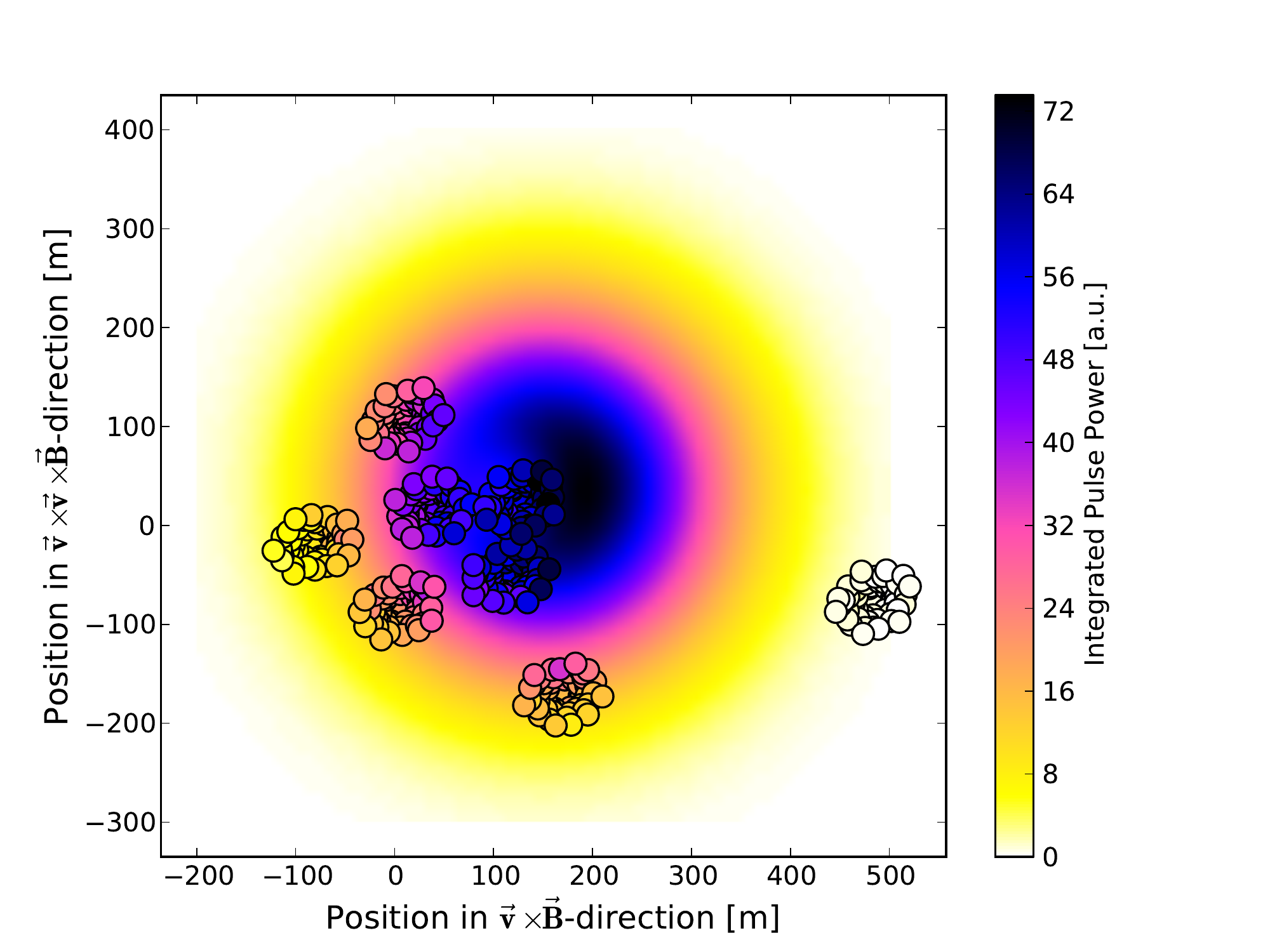}
\includegraphics[width=0.47\textwidth]{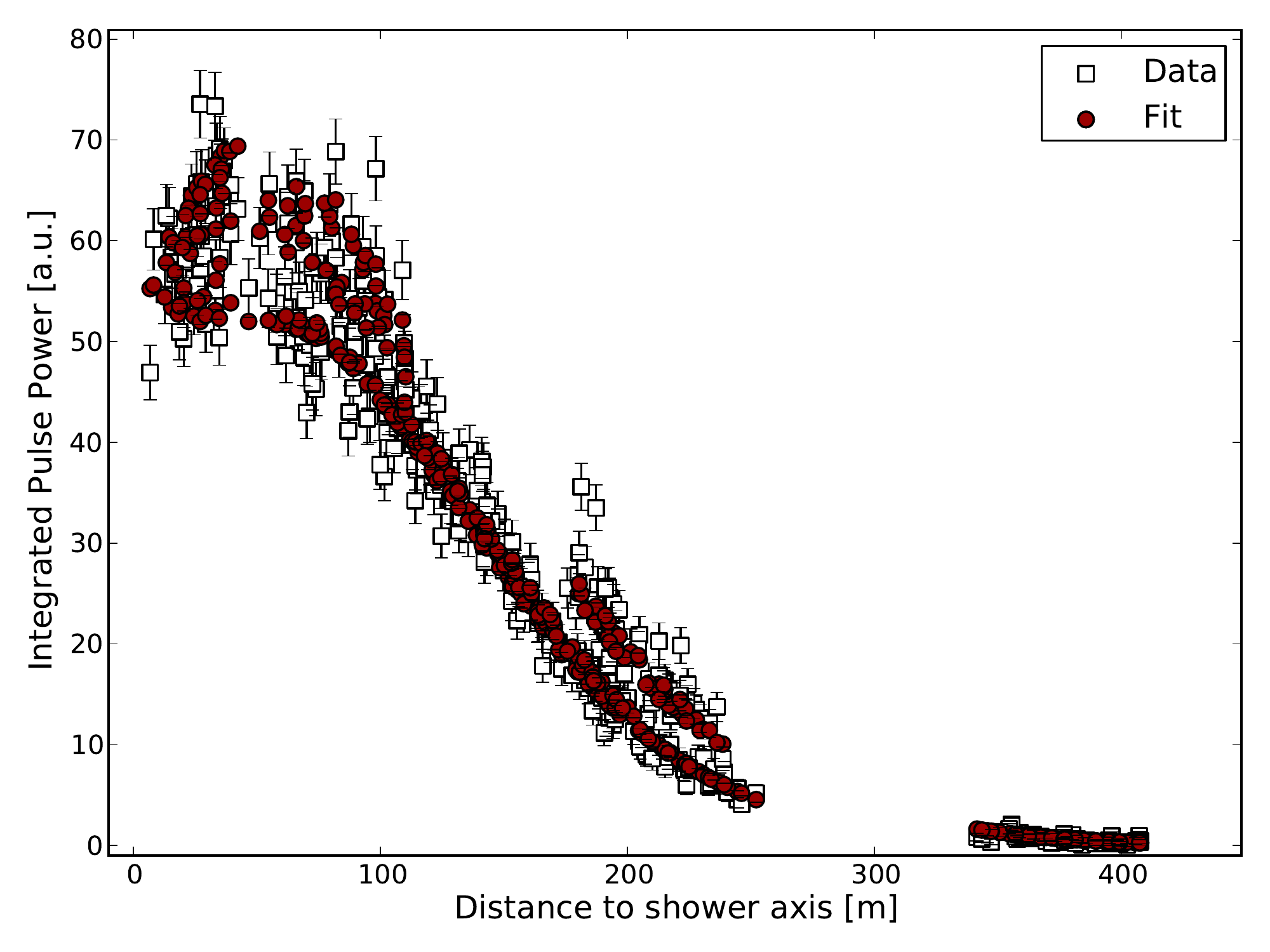}
\caption{Air showers as measured with LOFAR with the best fits to the data (equation (\ref{eq:fit})). Left: Pattern projected into the shower plane. The circles indicate the measurements, the background indicates the fit. The integrated total pulse power is encoded in color. Right: Pulse power as a function of the distance to the shower axis. The open black squares indicate the measurements, the full red circles show the fit to the data. }
\label{fig:ldf_fit}
\end{center}
\end{figure}

Function (\ref{eq:fit}) was tested with a set of LOFAR events. The events were reconstructed with the standard LOFAR reconstruction software \cite{Schellart2013}, delivering integrated pulse powers per antenna. The pulse powers are calibrated relatively with respect to each other. There is no absolute calibration (yet). The events were fitted with the free parameters $A_+, X_c, Y_c, \sigma_+, x_-,$ and $C_0$ in a restricted range. Figure \ref{fig:ldf_fit} shows two example events of this test set. The data and fit are shown on the left in the two dimensional shower plane (background fit, foreground data) and on the right as a function of distance to the shower axis, the \emph{classical} way to plot the signals from an air shower. Here, the fit is indicated in the red full circles.  As the LOFAR antennas are not placed on a regular grid, the fit is expected to be more challenging. However, the figure shows a very good agreement between data and fit. It especially shows that different locations in the shower plane with the same distance to the shower axis show different signals. This asymmetry is nicely represented by the fit. 

The achieved reduced $\chi^2$ for the fits are $1.3$ and $1.6$, respectively. This does not illustrate perfect fit quality for more than 300 data points. It should however be taken into account that the current uncertainties as shown in figure \ref{fig:ldf_fit} only represent the influence of the noise on the measurement. Instrumental effects are not included, therefore the uncertainties are likely to be underestimated. Further studies will show, what resolutions can be achieved for different shower parameters for the whole LOFAR data set. 

\section{Conclusions and Outlook}
We present a parameterization for the signal distribution of the radio emission from air showers at ground level. All parameters can (within expected experimental uncertainties) be reduced to physical parameters, namely the energy $E$ of the air shower, the depth of the shower maximum $X_{\mathrm{max}}$, the position of the shower axis $(X,Y)$, and the arrival direction $(\phi,\theta)$. This parameterization (eq.\ref{eq:full}) describes all air showers with an average uncertainty of 15\%. This includes all dependencies on arrival direction and value of $X_{\mathrm{max}}$. After once establishing the parameterization, no additional input from specific simulations is required.

In addition to the parameterization based on air shower parameters, we derive a function that is suitable to be used on experimental data (eq.\ref{eq:fit}).  The fit can essentially be reduced to four parameters, given that in experiments the arrival direction is estimated independently of the signal strength via timing. We exemplary show that the parameterization reproduces air showers as measured with LOFAR. This is the first analytic parameterization to do so. 

Due to the correlation of the fit parameters with energy and distance to $X_{\mathrm{max}}$, this parameterization can for LOFAR simplify and speed up the identification of $X_{\mathrm{max}}$ by reducing the parameter space for individual simulations for every air shower. In further investigations we will study methods to derive $X_{\mathrm{max}}$ based on this parametrization from measured data and explore the achievable resolution. 

If one wants to use the lateral distribution of the radio emission of air showers as an independent tool to determine all air shower characteristics, one needs to provide a sufficiently high number of  independent measurements of the signal strength. Experiments measuring the radio emission then need to be set-up accordingly. In oder to be able to use the discussed parametrization of the lateral distribution (eq.\ref{eq:fit}) in its most minimal form with the largest number of fixed parameters at least four measurements at different positions are needed. 

\section{Acknowledgments}
We would like to thank Christiaan Brinkerink for fruitful discussions concerning this model. This work was supported by the Foundation for Fundamental Research on Matter (FOM) and the Netherlands Organization for Scientific Research (NWO), VENI grant 639-041-130. We acknowledge funding from an Advanced Grant of the European Research Council under the European Unions Seventh Framework Program (FP/2007-2013) / ERC Grant Agreement n. 227610. Part of this work was also supported by grant number VH-NG-413 of the Helmholtz Association.

\appendix
\section{Fit parameters}
\begin{table}[h!]
\begin{tabular}{cc}
$C_i$&Fit value\\
\hline
$C_0$&$0.24\pm0.08$\\
$C_1$&$10^{-52.8\pm0.1}$ J $\cdot$ m$^{-2}\cdot$ eV$^{-2}$\\
$C_2$&$-7.88\pm0.09$ m\\
$C_3$&$28.58\pm0.06$ m\\
$C_4$&$1.98\pm0.03$ m\\
$C_5$&$-2.57\pm0.02$ m\\
$C_6$&$-54.9\pm0.7$ m\\
$C_7$&$0.44\pm0.01$ m $\cdot$ g$^{-1}\cdot$ cm$^2$\\
$C_8$&$-1.27\cdot10^{-4}\pm1\cdot10^{-6}$ m $\cdot$ (g$^{-1}\cdot$ cm$^2$)$^2$\\
$C_{9}$&$20.4\pm0.8$ m\\
$C_{10}$&$0.006\pm0.002$ m $\cdot$ g$^{-1}\cdot$ cm$^2$\\
$C_{11}$&$9.7\cdot10^{-5}\pm1 \cdot10^{-6}$ m $\cdot$ (g$^{-1}\cdot$ cm$^2$)$^2$\\
$C_{12,0}$&$107 \pm 4$ m\\
$C_{12,1}$&$-0.94\pm0.02$ m $\cdot$ g$^{-1}\cdot$ cm$^2$\\
$C_{12,2}$&$1.94\cdot10^{-3}\pm5\cdot10^{-5}$ m $\cdot$ (g$^{-1}\cdot$ cm$^2$)$^2$\\
$C_{12,3}$&$-1.5\cdot10^{-6} \pm 5\cdot10^{-8}$ m $\cdot$ (g$^{-1}\cdot$ cm$^2$)$^3$\\
$C_{12,4}$&$4.1\cdot10^{-10} \pm 2\cdot10^{-11}$ m $\cdot$ (g$^{-1}\cdot$ cm$^2$)$^4$\\
\end{tabular}
\caption{Parameters as obtained for equation (\ref{eq:full}). Note the parameters are calculated for the measurement situation of LOFAR. (Location at $\unit[5]{m}$ above sea level, geomagnetic field: $\unit[18.6]{\mu T}$ north and $\unit[45.6]{\mu T}$ downwards, bandpass filtered between $\unit[10-90]{MHz}$). }
\label{tab:parameters}
\end{table}

\bibliographystyle{model1-num-names}
\bibliography{ldf.bib}

\end{document}